\documentclass{article}
\usepackage{amsmath}

\usepackage{xcolor}
\usepackage{arxiv}
\usepackage[utf8]{inputenc} % allow utf-8 input
\usepackage[style=ieee]{biblatex}
\usepackage[T1]{fontenc}    % use 8-bit T1 fonts
\usepackage{hyperref}       % hyperlinks
\usepackage{url}            % simple URL typesetting
\usepackage{booktabs}       % professional-quality tables
\usepackage{amsfonts}       % blackboard math symbols
\usepackage{nicefrac}       % compact symbols for 1/2, etc.
\usepackage{microtype}      % microtypography
\usepackage{amsmath}
\usepackage{amssymb}
\usepackage{amsthm}
\usepackage{lipsum}
\usepackage{graphicx}
\usepackage{booktabs}
\usepackage{array}
\usepackage{multirow}
\usepackage{tabularx}
\usepackage{soul}
\usepackage{makecell}

\graphicspath{ {./images/} }

\title{Graph Analysis of Neuronal-Culture Connectivity Derived from a Reservoir-Computing Model}

\author{
 Ilya Auslender \\
  Department of Physics\\
  University of Trento\\
  Via Sommarive 14, Trento, TN, Italy, 38123 \\
  \texttt{ilya.auslender@unitn.it} \\
  \And
    Giorgio Letti \thanks{Current Address: Italian Institute of Technology, Via Morego 30, Genoa, Italy, 16163} \\
  Centre for Integrative Biology (CIBIO)\\
  University of Trento\\
  Via Sommarive 9, Trento, TN, Italy, 38123 \\
  \texttt{giorgio.letti@iit.it}
  \And
  Yasaman Heydari\\
  Department of Physics\\
  University of Trento\\
  Via Sommarive 14, Trento, TN, Italy, 38123 \\
  \texttt{seyedeh.heydari@unitn.it} \\
  \And
   Lorenzo Pavesi \\
  Department of Physics\\
  University of Trento\\
  Via Sommarive 14, Trento, TN, Italy, 38123 \\
  \texttt{lorenzo.pavesi@unitn.it}
}

\begin{document}
\maketitle

\begin{abstract}
Graph-theoretical analysis offers a principled framework for quantifying emergent dynamics in neuronal cultures. Here, we present an analytical pipeline for inferring network-level properties of in vitro cortical cultures from multichannel electrophysiological recordings. The approach builds on a recently proposed Reservoir Computing (RC) framework (Auslender \textit{et al.}, 2025), which enables direct extraction of an \textit{Intrinsic Connectivity Map} (ICM) from neural activity. We interpret the ICM as an effective adjacency matrix and apply graph-theoretic centrality measures to quantify node- and edge-level contributions to the culture’s collective dynamics. We systematically evaluate both local and global graph metrics and examine their relationships with experimentally measured activity features, including firing rates and network-level descriptors. To validate the inference procedure, we also simulate the experimental environment, enabling controlled benchmarking of the RC-derived connectivity against a known ground-truth adjacency matrix and assessment of model performance as a function of graph structure. Our results demonstrate statistically robust associations, of varying strength, between graph-theoretic measures and experimentally observed activity patterns. These findings additionally support the validity of the RC-based connectivity inference and establish a scalable, data-driven framework for functional network characterization in neuronal culture systems.
\end{abstract}

% keywords can be removed

\section{Introduction}
Studies of neuronal networks offer valuable insights into brain functionality from a systems-level perspective \cite{Kandel2013Principles,BassettSporns2017NetworkNeuroscience}. While the study of cellular functionality focuses on the mechanistic properties of individual biological units, a deeper understanding of network dynamics and morphology reveals how the structure and organization of interconnected neurons shape the overall function of specific brain regions and, ultimately, the intact brain \cite{Nelson2021,Bullmore2012}. Much like in transportation or social networks, identifying the role and significance of individual nodes can shed light on the functionality, or dysfunction, of the entire neural circuit \cite{BullmoreSporns2009Complex,RubinovSporns2010ComplexMeasures,Crossley2014HubsDisorders}.

The brain is increasingly conceptualized as a complex network composed of interconnected circuits and sub-circuits \cite{Schroeter2017,Sporns2014}, each supporting specialized functions that emerge from their physiological and morphological properties. These properties arise from circuit architecture as well as cellular and subcellular characteristics. A central component of such network-based approaches is the field of \textit{connectomics} \cite{Sporns2005Connectome,VanEssen2013HCP,Fornito2015Disorders}, which aims to infer functional and structural connectivity at multiple scales—from local microcircuits involving individual neurons or small neuronal assemblies to large-scale networks linking distinct brain regions.

Once a connectivity map is obtained, its structural organization can be rigorously characterized using \textit{graph theory} \cite{Diestel2017GraphTheory}. In this framework, the network is represented as a graph $\mathcal{G}(V,E)$, where $V$ denotes the set of vertices (nodes) and $E$ the set of edges (connections). This abstraction is broadly applicable: for example, nodes may correspond to cities linked by roads, or to neurons and neuronal populations connected by synapses in a connectome \cite{Newman2010Networks}. Graphs can be \emph{directed} or \emph{undirected} and \emph{weighted} or \emph{unweighted}, depending on whether directionality and connection strength are encoded. Graph-theoretic analysis enables quantitative assessment of network architecture \cite{Nelson2021,sporns2012simple}, including the identification of influential nodes (e.g., via centrality measures) and the detection of modular organization.

While graph theory provides a formal mathematical framework to analyze connectivity, \textit{connectomics} focuses on reconstructing that connectivity from experimental data. This typically involves neuroimaging or electrophysiological techniques---such as functional MRI (fMRI), electroencephalography (EEG), or invasive neural recordings---combined with computational methods to infer functional or structural links \cite{de2018connectivity,Smith2018}. The achievable resolution critically depends on the measurement modality: fMRI and EEG enable mapping connectivity between large-scale brain regions, whereas characterization of local microcircuits often requires invasive approaches capable of detecting activity at the level of individual neurons \cite{Logothetis2008FMRI,Buzsaki2004LargeScale,Lichtman2011Connectomics}. In basic neuroscience research, for example, \textit{in-vitro} neuronal preparations provide controlled environments that permit high-resolution recordings \cite{Paninski2018} and detailed investigation of local circuit dynamics.

Among invasive modalities, electrophysiological techniques play a central role in probing neuronal network activity \cite{Buzsaki2004LargeScale}. By recording extracellular voltage fluctuations, electrodes can capture spiking and synaptic events from neural populations \cite{Spira2013}. To isolate circuit mechanisms and reduce biological complexity, \textit{in-vitro} systems are frequently employed. A widely used platform is the microelectrode array (MEA), which consists of a grid of electrodes capable of simultaneously sampling extracellular signals from many neurons \cite{Obien2015}, thereby enabling the study of emergent network dynamics under highly controlled conditions \cite{Maccione2012,Chiappalone2008}.

This paper follows a recently developed model presented in \cite{AUSLENDER}, which leverages spatio-temporal electrophysiological data recorded via MEAs. Based on the \textit{Reservoir Computing Network} (RCN) framework, this model (RC-model) reconstructs the network’s connectivity map, described by the \textit{intrinsic connectivity matrix} (ICM) after training. Moreover, it can predict the network’s spatio-temporal responses to input stimuli. The model has demonstrated superior performance in connectivity inference compared to statistical methods such as Cross-Correlation (CC) \cite{Stetter2012} and Transfer Entropy (TE) \cite{Vicente2011}, as well as other machine learning approaches \cite{endo2021convolutional}. 

In this study, we investigate the ICM derived from the RC model, testing its usage as an adjacency matrix for graph-based post-processing \cite{Poli2015}. This approach extends the analysis of MEA recordings beyond local electrode activity measures (e.g., spike and burst rates) toward a network-level interpretation of connectivity. We focus on node centrality measures and their relationship to emergent measurable activity of neurons. By integrating these data with connectivity maps reconstructed through the RC model, we aim to deepen the interpretation of neuronal connectivity, relate it to observed activity, and provide complementary validation of the RC model through graph-theoretic measures.

\begin{figure}
    \centering
    \includegraphics[width=0.95\linewidth]{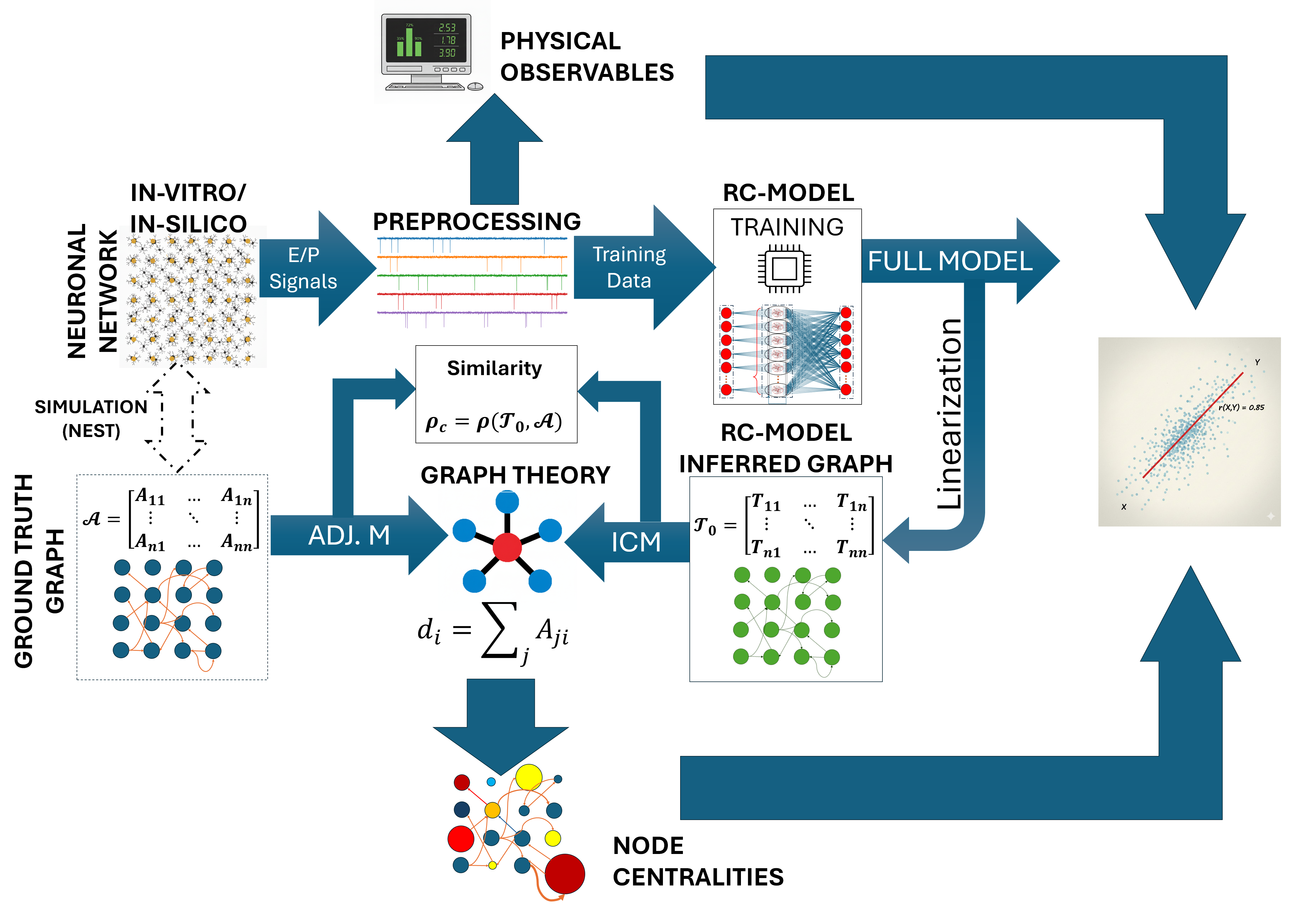}
    \caption{\textbf{Schematic pipeline of the present study}. Electrophysiological signals emerge from \textit{in-vitro} or \textit{in-silico} neuronal networks (simulated in NEST \cite{Gewaltig:NEST}), from which spatio-temporal activity patterns are preprocessed and quantitative observables (e.g., firing rate) are extracted. The preprocessed data are further transformed into training samples for the RC model, which infers the \textit{Intrinsic Connectivity Matrix} (ICM), denoted as $\mathcal{T}_0$ \cite{AUSLENDER}. Graph-theoretical analysis is then applied to the inferred connectivity, and the resulting graph measures (e.g., node centralities) are compared with experimentally measured activity observables. For instance, the node-degree metric is correlated with the measured firing rate at the electrode corresponding to that node. In parallel, structured networks generated \textit{in silico} provide a controlled framework for evaluating the performance of the RC model against known ground-truth graph properties of their adjacency matrices (ADJ. M), denoted as $\mathcal{A}$. In particular, the similarity $\rho_C$ between the inferred connectivity matrix $\mathcal{T}_0$ and the ground-truth adjacency matrix $\mathcal{A}$ is evaluated. This approach also enables the assessment of the relationship between structural network features and the resulting emergent dynamics.}
    \label{fig:scheme}
\end{figure}

\section{Materials and Methods}
\label{sec:methods}
\subsection{Methodological Framework}

We aim to relate network topology to emergent activity in multichannel electrophysiological recordings, moving beyond purely local interpretations. Let $n$ denote the number of nodes. We define two variable domains: the experimental domain,
\[
\mathbf{x}^{(p)} \in \mathbb{R}^{n},
\]
whose components represent the value of an observable $p$ at each node, and the graph-theoretical domain,
\[
\mathbf{v}^{(c)} \in \mathbb{R}^{n},
\]
whose components correspond to a centrality measure $c$ computed on the network $\mathcal{G}(V,E)$.

We investigate an implicit mapping
\begin{equation}
F^{(c,p)} : \mathbf{v}^{(c)} \rightarrow \mathbf{x}^{(p)},
\end{equation}
thereby testing whether node-level topological descriptors systematically explain variance in experimentally measured activity. Distinct selections of $(c,p)$ instantiating a specific structure–function hypotheses, enabling a quantitative assessment of topology–dynamics coupling. In this study, we focus primarily on the linear Pearson correlation coefficient $r_{VX}$ \cite{Pearson1895Regression} and explore candidate analytical relationships that may underlie the observed associations.

For the purposes of this study, we define the domain of $\mathbf{v}$ as the set of independent variables (IV), and the domain of $\mathbf{x}$ as the set of dependent variables (DV).

\subsection{Background on the Reservoir-Computing (RC) Model}
Since a comprehensive introduction to the RC model is given in \cite{AUSLENDER}, we summarize below only the essential background for the present analysis. A more detailed and rigorous description can be found therein. In that work, we demonstrated that the model can infer connectivity maps and predict neuronal network dynamics from experimental electrophysiological recordings of mouse cortical cells acquired with a microelectrode array (MEA). After standard preprocessing steps, including filtering and spike detection, the signals were transformed into multichannel instantaneous spike rate (ISR) sequences, from which short episodes of network bursting were extracted. These episodes were then used to train an artificial neural network (ANN) with a reservoir computing (RC) architecture, enabling it to learn the synaptic transmission function underlying the rate-coded sequences. The model assumes that information in the network is encoded in spike rates and transmitted according to the functional relationships learned from the experimental data.

In this framework, the real neuronal network is represented by a model network whose nodes correspond to the MEA electrodes. Since each electrode samples activity from an extracellular region that may encompass multiple neurons, the resulting representation corresponds to a macroscopic network, where nodes reflect neuronal circuits rather than individual neurons. At its core, the RC model incorporates a \textit{reservoir} layer that performs high-dimensional nonlinear transformations with leaky memory through recurrent connections. This stage reflects the inner circuitry of a node and is not modeled explicitly, but rather approximated by a randomly interconnected (with constraints) nonlinear micro-network. Finally, each time step of the sequence is captured through a linear regression with LASSO regularization, trained on the experimental ISR sequences. The model is trained such that the loss $\mathcal{L}$ between the predicted trace $\tilde{\mathbf{Y}}$ and the experimental trace $\mathbf{Y}$: $\mathcal{L} \left(\tilde{\mathbf{Y}},\mathbf{Y} \right)$ is minimal. Consequently, the trained model, given an initial state of the network state $\mathbf{y}[0]$ can predict the next time steps by applying:

\begin{equation}
\label{eq: model operator}
    \tilde{\mathbf{y}}[n+1] = \hat{\mathcal{F}} \Big\{ \tilde{\mathbf{y}}[n] \Big\}
\end{equation}
Here, $\hat{\mathcal{F}}$ denotes the operator of the trained model that advances the network state $\mathbf{y}[n]$ from discrete time step $n$ to $n+1$.

To this end, we tested the model by evaluating its predictions in response to an initial stimulation $\mathbf{y}[0]$, and compared them with the actual responses of the culture from which the training data were obtained. The results show that, within a certain degree of accuracy, the model is able to predict the spatio-temporal dynamics of neuronal activity given a specific input (or stimulus), both \textit{in silico} and \textit{in vitro} (i.e., in real living neuronal cultures).

Another outcome of the model was the inference of the network's connectivity map. This was achieved by linearizing the operator in Eq.\eqref{eq: model operator} and removing the influence of previous time steps on the connectivity. We therefore defined the resulting matrix as the \textit{Intrinsic Connectivity Matrix} (ICM), denoted by $\mathcal{T}_0$, as it represents the basal connections that are independent of the network state at any time. The model, however, updates the connectivity at each time step based on the current state and the history of previous states, modulated by the leakage (memory) parameter $\alpha$. We showed that, if the system remains in the linear regime for a few time steps (small-signal approximation), the connectivity can be represented by higher-order connectivity matrices, $\mathcal{T}_1$, $\mathcal{T}_2$, …, which couple the current state at time step $n$ to states at earlier time steps $n-2$, $n-3$, …

\subsection{Utilization of the RC-Model In This Work}

%$\mathcal{W}\text{in}$, $\mathcal{W}\text{res}$, and $\hat{\mathcal{S}}$

Preprocessed data from both simulated and experimental recordings were used to train the RC model. The training and validation were subjected to a root-mean-square-error (RMSE) loss function. This procedure was repeated seven times per network with different random initializations of the fixed weight matrices, where each initialization satisfied the constraints defined in Ref.~\cite{AUSLENDER}. The training and validation repetitions were performed to assess the stability of the model, particularly in the derivation of the Intrinsic Connectivity Map (ICM). To quantify this stability, we introduce a connectivity confidence measure, which characterizes the consistency of inferred connectivity across model initializations:
\begin{equation}
\label{eq: conn. conf.}
    \Gamma = 1-\frac{\max(\sigma_{\mathcal{T}_0})}{\max(\mu_{|\mathcal{T}_0|})}
\end{equation}
where $\mu_{|\mathcal{T}_0|}$ denotes the mean of each element of matrix $|\mathcal{T}_0|$ (under absolute value) and $\sigma_{\mathcal{T}_0}$ denotes the standard deviation of each element, computed across the seven repetitions.

For the data analyzed in this study, the reservoir dimension multiplier $m$ was set to a fixed value of 5, representing a practical compromise between representational richness and computational cost. The memory parameter $\alpha$ was optimized over the confidence $\Gamma$ within 0.1 to 0.9 range (refer to \cite{AUSLENDER} for details about the parameters). 

For both experimental recordings and simulations, the RC model was trained on spontaneous activity data. We additionally evaluated training on evoked responses; however, this did not yield qualitatively different results. Notably, the network inferred via the ICM was systematically smaller under evoked-response training, likely reflecting the more spatially localized nature of stimulus-driven activity.

\subsection{Network Activity Observables: The Dependent Variables}
We focus on the some common measures reported in many MEA measurements studies, e.g., \cite{Wagenaar2006,Obien2015}. 
\subsubsection{Node- (Electrode-) Based Measures}
\begin{enumerate}
    \item \textbf{Average Firing Rate (AFR)} (spike rate): Spikes constitute the fundamental units of neuronal communication. They manifest as rapid transient changes in membrane potential caused by ionic fluxes across the cell membrane, known as action potentials (APs). In MEA recordings, APs are detected extracellularly by electrodes positioned in the vicinity of individual neurons or small neuronal populations. Following spike detection from the raw voltage traces, the AFR is computed as the mean spike count per unit time over the duration of the recording \cite{Obien2015}.
    \item \textbf{Average Burst Rate (ABR)}: Bursts are episodes of high-frequency spike trains, composed of multiple spikes separated by short inter-spike intervals (ISIs), typically reflecting transient periods of enhanced excitability and coordinated network activity. They are commonly identified using threshold-based criteria on ISI or adaptive burst-detection algorithms \cite{pasquale2010self,chiappalone2005burst} applied to spike trains. The ABR is defined as the mean number of detected bursts per unit time over the recording duration.
    \item \textbf{Inter-Spike-Interval (ISI) Distribution Measures}: In neuronal networks, spike trains typically exhibit a bimodal temporal structure, characterized by short ISIs associated with intra-burst firing and longer intervals corresponding to inter-burst periods (IBIs) \cite{Wagenaar2006}. Given a sufficiently large number of detected spikes, an ISI histogram can be constructed to approximate the empirical ISI probability density function \cite{pasquale2010self}. From this distribution, several statistical measures can be extracted to characterize population spiking dynamics. In this work, we estimate the statistical mean of the ISI distribution, $\mu_{\mathrm{ISI}}$, as a measure of the characteristic spiking timescale of the ISI distribution. In addition, we compute the standard deviation of the ISI distribution, $\sigma_{\mathrm{ISI}}$, as a measure of temporal dispersion in the population spiking dynamics.
\end{enumerate}

\subsubsection{Overall Activity Measures}
Here, we define overall metrics to characterize the global activity of the culture, namely the mean firing rate (MFR) and the mean bursting rate (MBR). These quantities are computed as the respective node-level measures averaged over the total number $N$ of active nodes (electrodes) in the network, thereby providing aggregate descriptors of overall spiking and bursting dynamics.

\subsection{Network Graph Measures: The Independent Variables}
Consider the general case of a \textit{directed} and \textit{weighted} graph $\mathcal{G}(V,E)$, comprising both excitatory and inhibitory edges (i.e., positive and negative weights), and described by an adjacency matrix $\mathcal{A} = A_{i,j}$, where $A_{i,j}$ denotes the weight of the edge directed from node $j$ to node $i$. This directionality convention is adopted to remain consistent with the derivation of the ICM \cite{AUSLENDER}. Using graph-theoretic analysis, the importance or influence of each node within the network can be quantified through various centrality measures. 

\subsubsection{Node Centrality Measures}
In general, centrality metrics assign a numerical score to each node based on its structural position in the graph, capturing different aspects of topological relevance—such as the number and strength of its connections (degree/strength centrality), its role in mediating information flow along shortest paths (betweenness centrality), or its influence within the global connectivity structure (eigenvector- or spectral-based centralities). Each measure therefore probes a distinct structural notion of node importance in the directed and weighted network. In this work, we focus on a subset of fundamental centrality measures that we found to be most relevant to the characterization of neuronal dynamics \cite{Fletcher}.
\begin{enumerate}
    \item \textbf{In- and Out-Degree}: In-degree and out-degree quantify the total weight of a node’s incoming and outgoing connections, respectively, and provide a first-order estimate of information flow through the node. Because the network comprises both excitatory and inhibitory edges, we distinguish between signed (effective) and magnitude-based degree measures. Specifically, we define
    \begin{equation}
    \label{eq:degrees}
    \begin{aligned}
    d_{\text{in,eff}}^{(i)} &= \sum_{j} A_{ji}, \\
    d_{\text{in,abs}}^{(i)} &= \sum_{j} \left| A_{ji} \right|, \\
    d_{\text{out}}^{(i)}    &= \sum_{j}  A_{ij} .
    \end{aligned}
    \end{equation}
    Here, $d_{\mathrm{in,eff}}^{(i)}$ denotes the \emph{effective} in-degree of node $i$, in which inhibitory (negative) inputs reduce the net centrality contribution. In contrast, $d_{\mathrm{in,abs}}^{(i)}$ represents the \emph{absolute} in-degree, accounting only for the magnitudes of the incoming weights, irrespective of their sign. This distinction was introduced to capture, at first order, the influence of inhibition on node activity. 

    For the out-degree, $d_{\mathrm{out}}^{(i)}$, the direct first-order effect of a node on its own activity is generally negligible. Therefore, we employ the standard definition of out-degree, while preserving the signs of the outgoing connections.
    \item \textbf{Katz Centrality}: The Katz centrality \cite{katz1953new}, $C_K$, generalizes degree centrality by incorporating contributions from indirect (multi-step) paths in addition to immediate neighbors. By attenuating the influence of distant nodes through a damping factor, it accounts for both local and global connectivity. Formally, it is defined as:
    \begin{equation}
    \label{eq: Katz}
        C_{K,i} = \sum_j \Big\{ (\mathcal{I} - \alpha_K\mathcal{A})^{-1} \Big \}_{ji}
    \end{equation}
    where $C_{K,i}$ is the Katz score of node $i$; $\mathcal{I}$ is the identity matrix; $\mathcal{A}$ is the adjacency matrix; $\alpha_K$ is the distance cost factor. The expression quantifies the cumulative influence on node $i$ after infinitely many steps, each attenuated by $\alpha_K$: $\sum_n\left(\alpha_K \mathcal{A}\right)^n$, which converges to Eq.~\ref{eq: Katz} for $\alpha \leq 1/\varrho(\mathcal{A})$, where $\varrho(\mathcal{A})$ is the spectral radius of $\mathcal{A}$. The parameter $\alpha_K$ controls the influence of distant nodes: a contribution from node $j$ to node $i$ that passes through $n$ intermediate nodes (i.e., along a path of length $k+1$) is factorized by $\alpha_K^n$. This parameter plays a role that is partially analogous to the memory parameter $\alpha$ introduced in the RC model \cite{AUSLENDER}, as it scales, at each time step, the influence of the preceding network state on the current one. However, since Katz centrality does not involve the nonlinear reservoir dynamics, these parameters generally have different effects.

    \item \textbf{Eigenvector Centrality}: Eigenvector centrality \cite{Bonacich1972,Bonacich1987} quantifies a node’s influence by accounting not only for the number and strength of its connections, but also for the importance of the nodes to which it is connected. The measure is defined recursively: a node attains high centrality if it is connected to other highly central nodes. 

    Formally, eigenvector centrality corresponds to the components of the leading eigenvector of the adjacency matrix $\mathcal{A}$, satisfying
    \begin{equation}
    \label{eq:EV cent}
        \mathcal{A}\mathbf{c} = \lambda_{\max}\mathbf{c},
    \end{equation}
    where $\lambda_{\max}$ is the dominant eigenvalue and $\mathbf{c}$ is the centrality vector. In directed and weighted networks, this definition captures global structural influence, reflecting how strongly a node participates in the principal connectivity mode of the graph.
     \item \textbf{PageRank Centrality}: PageRank centrality \cite{Page1998} extends eigenvector-based importance measures to directed networks by incorporating a stochastic normalization and a damping mechanism. A node is considered important if it receives links from other important nodes, with contributions weighted by the out-degree of the source nodes. In contrast to standard eigenvector centrality, PageRank mitigates dominance by highly connected hubs and ensures well-defined scores even in the presence of sinks or disconnected components.

    Formally, the PageRank vector $\mathbf{p}$ satisfies
    \begin{equation}
    \label{eq:page rank}
        \mathbf{p} = \alpha \mathbf{P} \mathbf{p} + (1-\alpha_p)\mathbf{u},
    \end{equation}
    where $\mathbf{P}$ is the column-stochastic transition matrix derived from the adjacency matrix $\mathcal{A}$, $\alpha_p \in (0,1)$ is the damping factor controlling the contribution of network structure, and $\mathbf{u}$ is a personalization vector (typically uniform). In directed and weighted networks, PageRank captures a global notion of influence based on the steady-state distribution of a random walk over the graph.

\end{enumerate}
\subsubsection{Global Graph Measures}

\begin{enumerate}
    
    \item \textbf{Inhibitory-Excitatory Balance}: The RC model demonstrated good performance in distinguishing and predicting inhibitory versus excitatory weights in the network. Therefore, we can define the inhibitory–excitatory ratio coefficient as:
    
    \begin{equation}
    \label{eq: inh-exc}
        \eta = \frac{\sum_{w<0}|w|}{\sum_{w>0} w} = \frac{W_{\text{inh}}}{W_{\text{exc}}},
    \end{equation}
    where $w$ is a weight of an edge $\mathcal{A}_{ij}$ for any $i,j$ within the adjacency matrix, $W_{\text{inh,exc}}$ is the total inhibitory or excitatory weight of the network.
    
    \item \textbf{Spectral Radius}: The spectral radius of the adjacency matrix is defined as the largest eigenvalue in absolute value,   namely
    \begin{equation}
        \varrho(\mathcal{A}) = \max_{k} |\lambda_k|,
    \end{equation}
    where $\{\lambda_k\}$ are the eigenvalues of $\mathcal{A}$. In directed and weighted networks, the spectral radius reflects the overall strength of connectivity and is closely related to feedback structure and the amplification potential of recurrent pathways.

    In addition to the weighted spectral radius, we define an unweighted and unsigned version
    \begin{equation}
    \label{eq:Spectral R}
        \varrho_0 = \varrho(\widetilde{\mathcal{A}}),
    \end{equation}
    where $\widetilde{\mathcal{A}}$ is the binary adjacency matrix obtained from $\mathcal{A}$ by
    \begin{equation}
        \widetilde{A}_{ij} =
        \begin{cases}
            1, & A_{ij} \neq 0, \\
            0, & A_{ij} = 0.
        \end{cases}
    \end{equation}
    This metric isolates purely topological contributions by removing weight magnitude and sign, thereby providing a structural measure of network complexity independent of edge strength.
    \item \textbf{Degree Density}: We define the degree density as
    \begin{equation}
    \label{eq:degree density}
        \delta_w^{(\text{exc})} = \frac{W_{\text{exc}}}{N}, \quad \delta_w^{(\text{inh})} = \frac{W_{\text{inh}}}{N}
    \end{equation}
    where $W_{\text{exc}}$ and $W_{\text{inh}}$ denote the total excitatory and inhibitory weight in the network, respectively, and $N$ is the number of active nodes. This quantity represents the average excitatory or inhibitory connectivity weight per node, thereby providing a normalized measure of the overall strength of synaptic drive in the culture.

\end{enumerate}
\subsection{ICM as an Adjacency Matrix}
In this work, we represent the neuronal culture as a graph in which nodes correspond to measurement sites (electrodes) and edges encode pairwise functional interactions. The Intrinsic Connectivity Map (ICM), denoted $\mathcal{T}_0$, is inferred via the RC model and assigns directed, signed weights to node pairs (positive for excitatory influence, negative for inhibitory influence). Accordingly, the system is described as a directed, weighted network, represented by a generally non-symmetric matrix.

Formally, the adjacency matrix $\mathcal{A}$ represents the underlying structural connectivity of the network, whereas the ICM is obtained by learning functional dependencies directly from observed activity. A central working hypothesis of this study is to treat the ICM as an effective adjacency matrix and to apply graph-theoretic measures to $\mathcal{T}_0$ in order to characterize structural properties of the culture.

This assumption is supported by previous validation \cite{AUSLENDER}, where simulations performed with NEST demonstrated that the inferred ICM exhibits, on average, high accuracy when benchmarked against the ground-truth adjacency matrix. Notably, this agreement was achieved despite the linearization of the RC model used for inference. Since $\mathcal{T}_0$ captures the leading-order (first-order) interaction structure of the system, we adopt the approximation
\begin{equation}
    \mathcal{T}_0 \simeq \mathcal{A},
\end{equation}
interpreting the ICM as an effective structural proxy for subsequent graph-theoretic analysis.

\subsubsection{ICM Preprocessing for Graph Analysis}
The ICM, obtained from the RC model following the procedure detailed in Ref.~\cite{AUSLENDER}, was normalized on an experiment-wise basis; that is, each matrix corresponding to a distinct experiment (or simulation) was processed independently. Because the RC model is trained on instantaneous spike-rate time series with comparable physical ranges, the inferred ICM weights typically occupy similar numerical scales across experiments. For each matrix, normalization was performed by rescaling all edge weights with respect to the maximum absolute weight within that matrix, thereby preserving the relative magnitudes between excitatory and inhibitory interactions. Consequently, the normalization step does not introduce substantial inter-experiment distortions, but rather ensures consistent scaling.

Specifically, if $\mathbf{w}$ denotes the set of all edge weights in a given ICM, each weight was transformed according to
\begin{equation}
\label{eq:weight norm}
    w_i' = \frac{w_i}{\max\left(|\mathbf{w}|\right)}.
\end{equation}
Consequently, the weight with the largest magnitude is mapped to either $+1$ or $-1$ (depending on its sign), and all remaining weights are scaled proportionally relative to this reference value.

This normalization facilitated the definition of a threshold for discriminating between genuine edges and negligible weights arising from numerical noise in the inference process. The threshold was empirically estimated using F1-score analysis (see Sec.~\ref{sec:results conn}) and set to $w'_{th} = 0.2$ in the normalized scale.

Accordingly, the thresholded weights were modified as following:
\begin{equation}
\label{eq:ICM threshold}
    \tilde{w}'_i =
    \begin{cases}
        w_i', & |w_i'| \geq w'_{th}, \\
        0,     & |w_i'| < w'_{th}.
    \end{cases}
\end{equation}
This procedure yields a sparsified connectivity matrix in which weak, noise-dominated interactions are removed while preserving the dominant excitatory and inhibitory structure of the inferred network.

\subsection{Data}
\subsubsection{MEA Measurements and Data Pre-Processing}
\label{sec: MEA experiments}
The experimental data used in this study were obtained from multielectrode array (MEA) recordings using 60 electrodes, collected from cortical neurons cultured from E17 embryonic mice. Recordings included both spontaneous activity and optogenetically evoked responses induced by light stimulation, using digital light processor (DLP) system [], providing 300 stimuli train of different spatial patterns at a frequency of 2 Hz. Raw signals underwent standard preprocessing, including filtering, spike detection, and burst identification, using custom MATLAB routines. All procedures for culture preparation, MEA recording, and data preprocessing followed the protocols described in Ref.~\cite{AUSLENDER}, where full methodological details are provided in the supplementary materials.

\subsubsection{\textit{In-Silico} Simulation}
We developed an \textit{in-silico} model using NEST \cite{Gewaltig:NEST}, designed to emulate microelectrode array (MEA) recordings and to benchmark the RC-based inference against structurally known ground-truth networks. The simulated network consisted of point-process neurons governed by the Izhikevich model \cite{izhikevich2003simple}, parameterized to reproduce regular-spiking dynamics \cite{izhikevich2004model}. 

A fixed number of neurons were grouped into distinct populations, each population representing a single node in the effective network and serving as an analogue of a MEA electrode sampling a localized neuronal assembly. Neurons within each population were fully connected, forming densely interconnected local hubs. In contrast, populations were sparsely and randomly interconnected via \textit{inter-population} synapses. Based on this construction, we derived a corresponding adjacency matrix (i.e., node-to-node adjacency) describing the effective connections between the defined populations. This effective connectivity is closely related to the ground-truth \textit{Structural Connectivity} of the network, as it accounts for inter-population connections while incorporating the effects of intra-population connectivity.

This hierarchical architecture was chosen to reproduce the emergence of highly interconnected hubs \cite{antonello2022self}, where each population represents a compact neuronal assembly surrounding a MEA electrode from which action potentials are recorded. Inter-population connections, in turn, emulate longer-range functional interactions between spatially separated assemblies.

A detailed description of the \textit{in-silico} model is provided in the supplementary materials of Ref.~\cite{AUSLENDER}.

\section{Results}
The analysis in this section proceeds along two complementary directions. The first investigates how ground-truth global graph measures, used to define the reference networks that generate the \textit{in-silico} neuronal activity, influence the validity, robustness, and interpretability of the connectivity reconstructed by the RC model via the ICM. In this context, we evaluate how specific structural properties impact inference accuracy and the recovery of known network characteristics.

The second direction examines how graph-theoretic properties of the connectivity, whether the ground-truth adjacency in simulations or the ICM inferred from experimental recordings, relate to measurable activity features of the culture. The goal is to determine to what extent structural descriptors can explain or predict the observed network dynamics. In this part, we analyze both global metrics and local node-level centralities alongside their corresponding activity observables.

For the following results we produced 40 \textit{in-silico} networks and 170 experiments taken from 54 different \textit{in-vitro} cultures.
\subsection{Definition of the Variable Space}
Table \ref{tab:variables} lists the variables examined in this study and categorizes them.
\begin{table}[]
\label{tab:variables}
    \centering
    \caption{\textbf{Summary of the variables examined in this study.} Graph-theoretical parameters are categorized as \textit{independent variables} (IV), whereas experimentally measured physical observables are defined as \textit{dependent variables} (DV).}
    \begin{tabularx}{\textwidth}{|m{1.6cm}| m{3.5cm} m{1.4cm} m{1.8cm} X|}
        \hline
        \centering Type & Variable/Parameter & Denotation & Definition & Significance\\
        \hline
        \multirow{3}{1.6cm}{Global Graph Variables (IV)}
         & Unweighted Spectral Radius & $\varrho_0$ & Eq.\eqref{eq:Spectral R} & Quantifies the complexity of the circuitry.\\
         \cline{2-5}
         & Excitatory/ Inhibitory Degree Density & $\delta_w^{(exc,inh)}$ & Eq.\eqref{eq:degree density} & \small Describes the average excitatory/inhibitory weight distribution per node. \\
         \cline{2-5}
         & Inhibitory to Excitatory Weight Ratio & $\eta$ & Eq. \eqref{eq: inh-exc} & \footnotesize Shows the balance between inhibition and excitation ($\eta = 1$ means balance between excitatory and inhibitory circuits; $\eta=0$: Fully excitatory network; $\eta > 1$: Network is predominately inhibitory).\\
         \hline

         \multirow{3}{1.6cm}{RC Model Performance Metrics (DV)} & \footnotesize Graph Similarity (ground truth vs. ICM). & $\rho_C$ & \cite{AUSLENDER} & \footnotesize Describes how close the ICM is associated with the structural connectivity of the network. This metric is available within simulations.\\
         \cline{2-5}
         & ICM Confidence & $\Gamma$ & Eq. \eqref{eq: conn. conf.} & \small Describes the stability of the ICM with respect to the initial parametrization of the RC model.\\ \cline{2-5}
         & Training/ Validation Loss & $\mathcal{L}_{tr,val}$ & \cite{AUSLENDER} & \footnotesize Quantifies the training performance of the RC model on the given spatio-temporal dataset of the tested culture, and assesses the extent to which the training data yield interpretable structure in the learned representation.\\ \hline

         \multirow{2}{1.6cm}{Network Mean Observables (DV)} & \small Mean Firing Rate (MFR) & $\overline{\langle r_f \rangle }$ & \scriptsize Average spike rate over nodes. & \small Describes the overall activity rate of the culture.\\ \cline{2-5}
         & \small Mean Bursting Rate (MBR) & $\overline{\langle r_b \rangle }$ & \scriptsize Average burst rate over nodes. & \small Describes the overall bursting tendency of the culture.\\ \hline

         \multirow{6}{1.6cm}{Nodes Centrality Measures (IV)} & Effective In-Degree & $d_{\mathrm{in,eff}}$ & Eq. \eqref{eq:degrees} &  \footnotesize Quantifies the total incoming synaptic weight of a node, explicitly accounting for the sign of each connection.\\ \cline{2-5}
         & Absolute In-Degree & $d_{\mathrm{in,abs}}$ & Eq. \eqref{eq:degrees} & \footnotesize Quantifies the total incoming synaptic weight of a node, without accounting for the sign of each connection. \\ \cline{2-5}
         & Out-Degree & $d_{\mathrm{out}}$ & Eq. \eqref{eq:degrees} & \small  Quantifies the total outgoing synaptic weight of a node, taking in account the sign of each connection. \\ \cline{2-5}
         & Katz Centrality & $C_K$ & Eq. \eqref{eq: Katz} &  Node importance accounting far interactions effects. \\ \cline{2-5}
         & Eigenvector Centrality & $C_{EV}$ & Eq. \eqref{eq:EV cent} &  \small Node importance determined recursively through interactions with the importance of other nodes in the network.\\ \cline{2-5}
         & PageRank Centrality & $C_{PR}$ & Eq. \eqref{eq:page rank} &  \small Node importance defined by a random-walk process, where influence is distributed through incoming links and normalized by the connectivity of the source nodes.\\ \hline

         \multirow{5}{1.6cm}{Node Observables (DV)} & \small Average Firing Rate (AFR) & $\langle r_f \rangle$ & \scriptsize Spike count per unit of time. & Node spiking activity. \\ \cline{2-5}
         & \small Average Bursting Rate (ABR) & $\langle r_b \rangle$ & \scriptsize Burst count per unit of time. &  Node bursting activity. \\ \cline{2-5}
         & \small Average Burst Duration (ABD) & $\langle t_B\rangle$ & \scriptsize Time difference between first and last spikes in a burst-- averaged over the recording.  & Average duration of bursting episode at a node. \\ \cline{2-5}
         & \small Characteristic ISI Timescale & $\mu_{\mathrm{ISI}}$ 
         & \scriptsize Statistical mean of the ISI distribution. 
         & \small Characteristic spiking timescale of the population. \\ \cline{2-5}
         & \small ISI Temporal Dispersion & $\sigma_{\mathrm{ISI}}$ 
         & \scriptsize Standard deviation of the ISI distribution. 
         & \small Quantifies temporal dispersion in population spiking dynamics. \\ \hline
    \end{tabularx}

    \label{tab:placeholder}
\end{table}

\subsection{Effect of Global Graph Properties on RC-model Performance}
\label{sec:results conn}

%\textit{In-silico} simulations in this work were used to provide ground-truth connectivity for evaluating graph-based measures and to assess the limitations of the RC model in inferring connectivity from known network structures. We performed 40 simulations of networks with varying sizes and configurations. 

In our previous study \cite{AUSLENDER}, we demonstrated the outstanding performance of the reservoir computing (RC) model in predicting the ground-truth structure of neuronal networks using NEST simulations. In the present work, we examine the conditions under which the linearization of the RC model, employed during ICM retrieval, provides an accurate and reliable prediction of the ground-truth structural connectivity.

\subsubsection{Effect of Inhibitory Strength on the Agreement Between the ICM and the Ground-Truth Adjacency Matrix}
Among the structural parameters that critically shape network dynamics, the overall level of inhibition plays a dominant role. It is quantified by the inhibitory–excitatory ratio $\eta$ defined in Eq.~\eqref{eq: inh-exc}. We find that this parameter directly modulates the agreement between the ICM and the ground-truth connectivity matrix, suggesting that the inhibitory–excitatory balance strongly shapes the strength of the nonlinear network dynamics and, consequently, impacts the accuracy of the linearized RC-based connectivity reconstruction.

We categorized the 40 networks into three groups according to their inhibitory–excitatory ratio:
\begin{itemize}
\item \textbf{Fully excitatory}: $\eta <0.1$, corresponding to negligible inhibitory weight.
\item \textbf{Low inhibition}: $0.1 \leq \eta < 0.5$.
\item \textbf{High inhibition}: $\eta \geq 0.5$.
\end{itemize}

We first evaluated the accuracy of connectivity prediction across all edges identified in each of the three groups. Both the ICM-derived and ground-truth connectivity matrices were normalized according to Eq.\ref{eq:weight norm}. Note that at this stage the ICM weights were not thresholded (Eq.~\ref{eq:ICM threshold}) in order to allow unrestricted, threshold-dependent evaluation of binary accuracy metrics.

\begin{figure}[htbp!]
    \centering
    \includegraphics[width=0.98\linewidth]{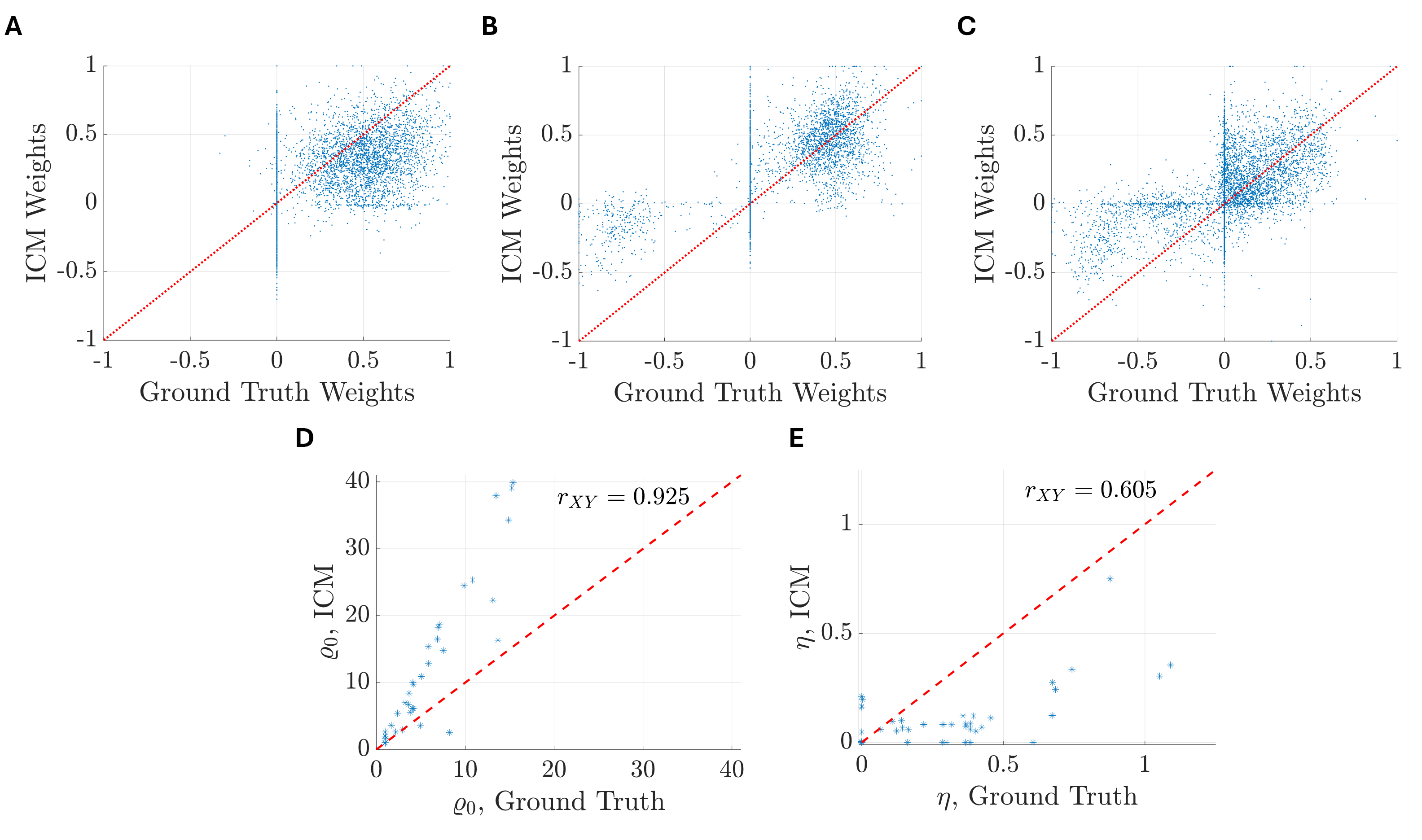}
    \caption{\textbf{Connectivity inference by the RC model via the \textit{Intrinsic Connectivity Matrix} (ICM)}. (A–C) Predicted vs.\ ground-truth structural weights for all 40 simulated networks, grouped by inhibitory–excitatory ratio: (A) fully excitatory, $\eta < 0.1$ (14\,720 edges; 9 networks); (B) low inhibition, $0.1 \leq \eta < 0.5$ (12\,928 edges; 23 networks); (C) high inhibition, $\eta \geq 0.5$ (17\,728 edges; 8 networks). For each network, weights were normalized by the maximum absolute synaptic weight. (D) ICM-predicted unweighted spectral radius $\varrho_0$ vs.\ ground truth. Pearson correlation $r_{XY}$ is indicated above. (E) ICM-predicted inhibitory–excitatory ratio $\eta$ vs.\ ground truth. Pearson correlation $r_{XY}$ is indicated above. In all figures, the red dashed lines indicate the unity-slope line, i.e., perfect prediction of the ground-truth values by the RC model.}

    \label{fig: weights prediction}
\end{figure}

Figures~\ref{fig: weights prediction}.A-C present the ICM-predicted edge weights as a function of their corresponding ground-truth values for each group. Because the number of excitatory and inhibitory connections differs substantially across networks and configurations, relying on a single performance metric may lead to biased or misleading conclusions. We therefore employed a set of complementary accuracy measures, each capturing a distinct aspect of prediction quality and collectively providing a robust and balanced evaluation of model performance.

\begin{itemize}
    \item \textbf{Receiver Operating Characteristic (ROC) - Area Under the Curve (AUC) of connection detection} \cite{Swets1988}: a threshold-independent metric assessing the ability to discriminate between existing and non-existing connections. ROC AUC is insensitive to class prevalence and is therefore well suited for evaluating binary connection detection under varying network sparsity.
    \item \textbf{Excitatory and inhibitory Precision-Recall (PR) AUC} \cite{Davis2006}: PR AUC, computed separately for excitatory and inhibitory connections, assessing the classification performance of the model in correctly detecting a connection of a given type in contrast to both the opposite synaptic type and non-existent connections. PR AUC is particularly informative in the presence of strong class imbalance and highlights performance on minority classes that may be underrepresented in a network.

    \item \textbf{Excitatory and inhibitory F1 score} \cite{Rijsbergen1979,Powers2011}: the harmonic mean of precision and recall, evaluated independently for excitatory and inhibitory edges. This metric provides an interpretable balance between false positives and false negatives at a specific operating point, which was selected using the optimal classification threshold.

    \item \textbf{Normalized Mean Weight Accuracy (NMWA)}: a continuous metric quantifying the average deviation of the predicted edge weights from their ground-truth values, normalized by the maximal possible deviation (equal to 2 after normalization),
    \begin{equation}
        \text{NMWA} = 1 - \frac{1}{2N}\sum_i \left| x_i - y_i \right| ,
    \end{equation}
    where $x_i$ and $y_i$ denote the normalized predicted and ground-truth weights, respectively. NMWA directly assesses weight reconstruction accuracy independently of thresholding.

    \item \textbf{Overall Pearson correlation}: the linear correlation coefficient between predicted and ground-truth edge weights, capturing global agreement in the relative ordering and sign of synaptic strengths across the entire dataset.

\end{itemize}

Table \ref{tab: ICM accuracy results} summarizes the results of these metrics for each of the network types.

\begin{table}[htbp!]
    \centering
    \caption{Results of the different metrics used to evaluate the accuracy of the ICM in reconstructing the ground-truth structural connectivity. The definitions of the reported metrics are provided above. Missing entries correspond to inhibitory connections in a full excitatory network. Although a small number of inhibitory connections are present, their influence on the network dynamics is marginal; consequently, their precise identification is both challenging and of limited relevance. For the F1 score, values in parentheses indicate the optimal threshold.}
    
    \begin{tabular}{|p{2.5cm}||m{1.1cm} m{1.1cm} m{1.1cm} m{1.1cm} m{1.1cm} m{1.1cm} m{1.15cm}|}
    \hline
        Network type &  ROC AUC & Exc. PRAUC  & Inh. PRAUC & F1 exc. (thr.) & F1 inh. (thr.) & NMWA & Ovr. Pearson \\

        \hline
         Full Excitatory, $\eta <0.1$, Fig. \ref{fig: weights prediction}.A & $0.851$	& $0.698$	& \centering --	& $0.658$ $(0.23)$	& \centering --	& $0.933$ & $0.58$\\
         \hline
         Low Inhibition, $0.1 \leq \eta  <0.5$, Fig. \ref{fig: weights prediction}.B & $0.922$	& $0.798$ &	$0.441$ & $0.796$ $(0.27)$ &	$0.482$ $(-0.17)$ & $0.95$ & $0.692$ \\
         \hline
         High Inhibition, $\eta \geq 0.5$, Fig. \ref{fig: weights prediction}.C & $0.759$  &	$0.501$	& $0.291$ &	$0.505$ $(0.17)$ &	$0.339$ $(-0.15)$ &	$0.946$ & $0.473$\\
         \hline
    \end{tabular}

    \label{tab: ICM accuracy results}
\end{table}

The accuracy measures discussed above, which compare the predicted weights to the ground-truth values, may not be sufficient to fully assess the quality of the reconstructed graph. Therefore, we also examined how higher-level structural properties are recovered. In particular, we evaluated the prediction of the network complexity, quantified by the unweighted spectral radius $\varrho_0$ (Eq.~\eqref{eq:Spectral R}), and the overall inhibitory-to-excitatory ratio $\eta$ (Eq.~\eqref{eq: inh-exc}), relative to their ground-truth values derived from the adjacency matrix. These quantities were considered because, unlike degree-based metrics, they are not necessarily linearly related to the individual edge weights and therefore provide a more global assessment of reconstruction quality. The prediction results for these measures are shown in Fig.~\ref{fig: weights prediction}D–E.

% ADDED
As shown in Fig. \ref{fig: weights prediction}A–C and Table \ref{tab: ICM accuracy results}, the level of inhibition significantly influences the agreement between ICM-derived estimates and the ground-truth structural connectivity. Notably, an intermediate inhibition regime yields maximal accuracy across all evaluated metrics, indicating the presence of an optimal operating point. In both the fully excitatory condition (Fig. \ref{fig: weights prediction}A) and the strongly inhibitory regime (Fig. \ref{fig: weights prediction}C), binary classification performance remains relatively high (as reflected by ROC AUC; Table \ref{tab: ICM accuracy results}); however, the linear correspondence between predicted and ground-truth weights is substantially degraded (overall Pearson correlation; Table \ref{tab: ICM accuracy results}). Furthermore, at high inhibition levels, the identification of excitatory versus inhibitory connections deteriorates, as evidenced by reduced PRAUC and F1 scores (Table \ref{tab: ICM accuracy results}). This trend is further supported by Fig. \ref{fig: weights prediction}E, which shows that balance prediction is optimal around $\eta \sim 0.1$ and deteriorates as the system moves away from this regime, reinforcing the existence of an optimal inhibition level. Network complexity, as measured by the unweighted spectral radius $\varrho_0$, exhibits an approximately proportional relationship with the ground-truth values (Fig. \ref{fig: weights prediction}D), but with a slope exceeding unity, reflecting a systematic overestimation. Given that the modeled network matches the ground truth in node count, this bias likely originates from the inference of additional intra-cluster connectivity during training.

\subsubsection{RC-Model Performance in Relation to Ground-Truth Global Graph Metrics}
The similarity between the ICM and the ground-truth connectivity alone does not fully characterize the performance of the RC model. Rather, it quantifies how accurately the ICM, retrieved after linearization of the model, reflects the underlying structural connectivity. A more direct assessment of model performance is provided by the training and validation losses, which indicate how well the model captures the observed network dynamics. In addition, since the present work focuses specifically on the connectivity properties encoded by the ICM, we assess its robustness through the confidence measure $\Gamma$ defined in Eq.~\eqref{eq: conn. conf.}.

\begin{figure}
    \centering
    \includegraphics[width=0.98\linewidth]{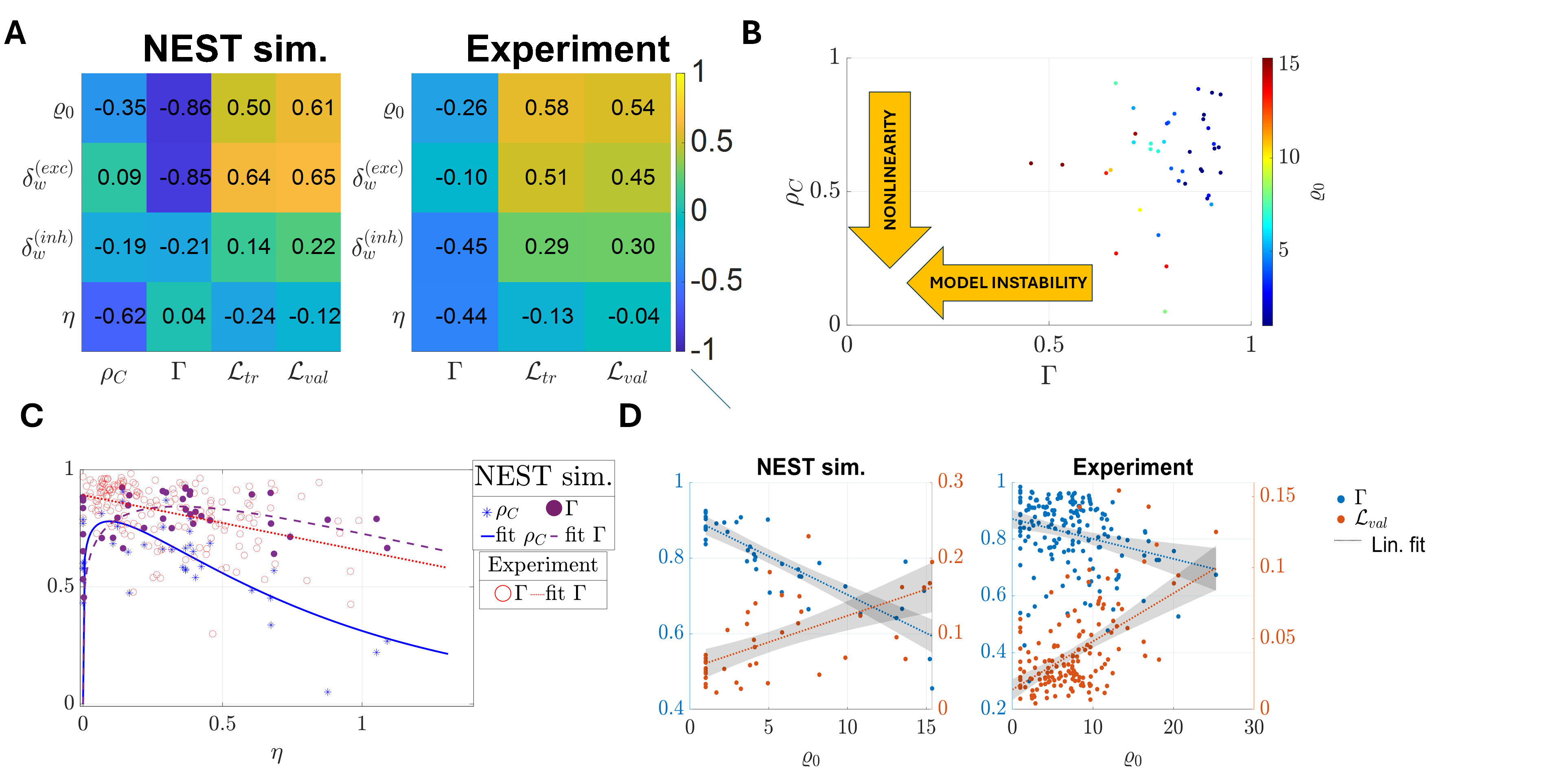}
    \caption{\textbf{Effect of global graph measures on RC-model performance}. (A) Pearson correlation map between ground-truth graph measures (vertical axis) and RC-model performance metrics (horizontal axis) for both NEST simulations and MEA experiments. (B) ICM confidence ($\Gamma$) does not necessarily correspond to the $\mathcal{T}_0 \leftrightarrow \mathcal{A}$ correlation ($\rho_C$). Increasing network complexity, quantified by the unweighted spectral radius $\varrho_0$, tends to induce stronger nonlinear dynamics, reducing $\rho_C$ and lowering $\Gamma$, which leads to less stable predictions. (C) Effect of the inhibitory level on RC-model performance. This parameter mainly affects connectivity prediction accuracy rather than prediction stability. A gamma-like distribution function $ax^p e^{-qx}$ is fitted, suggesting the existence of an optimal inhibition level. (D) Dependence of model performance on $\varrho_0$, evaluated through the confidence $\Gamma$ and the validation loss ($\mathcal{L}_{val}$), which is strongly correlated with the training loss ($\mathcal{L}_{tr}$; see (A)). A linear fit with 95\% confidence indicates decreasing model performance with increasing network complexity.}
    \label{fig:performance}
\end{figure}

% ADDED
Fig.~\ref{fig:performance} summarizes the dependence of RC model performance on global network properties. We observe that the spectral radius of the ground-truth connectivity matrix, $\varrho_0$, has a direct impact on model performance. Specifically, larger spectral radii are associated with reduced ICM confidence and increased training and validation losses. This trend is evident in the correlation map (Fig. \ref{fig:performance}A), which shows strong negative correlations between $\Gamma$ and $\varrho_0$, and positive correlations between $\Gamma$ and the losses $\mathcal{L}$.

This behavior can be attributed to the increased dynamical variability induced by stronger and more heterogeneous coupling across neuronal populations. Such complexity produces datasets that are more challenging to fit using linear regression, as employed in RC frameworks, and simultaneously degrades the accuracy of the linearization procedure underlying ICM retrieval.

To further illustrate that connectivity similarity $\rho_C$ and model stability (quantified by $\Gamma$) do not necessarily co-vary, we represent them in a “phase space” (Fig. \ref{fig:performance}B). The distribution shows that, for larger $\varrho_0$, the model is less likely to achieve high performance in either metric. This regime is consistent with increased nonlinearity (affecting $\rho_C$) and reduced learning stability (affecting $\Gamma$).

Model performance can also be linked to the inhibitory level $\eta$, consistent with the trends observed in the previous subsection on connectivity retrieval. Figure \ref{fig:performance}C illustrates the dependence of both $\Gamma$ and $\rho_C$ on $\eta$. As discussed above, an optimal inhibition regime is expected; accordingly, we approximate this dependence using a phenomenological gamma-like distribution function of the form $a x^p e^{-q x}$. This fit captures a steep rise at low inhibition ($\eta \sim 0$), followed by a maximum around $\eta \sim 0.1$–$0.2$ and a gradual decay at higher values.

Figure \ref{fig:performance}D further highlights the degradation in performance, quantified by $\Gamma$ and the validation loss $\mathcal{L}_{\text{val}}$, as a function of $\varrho_0$. A linear fit emphasizes the systematic decline in performance with increasing spectral radius.

\subsection{Relationship Between Global Graph Properties and Culture-Level Physical Observables}
\begin{figure}
    \centering
    \includegraphics[width=0.98\linewidth]{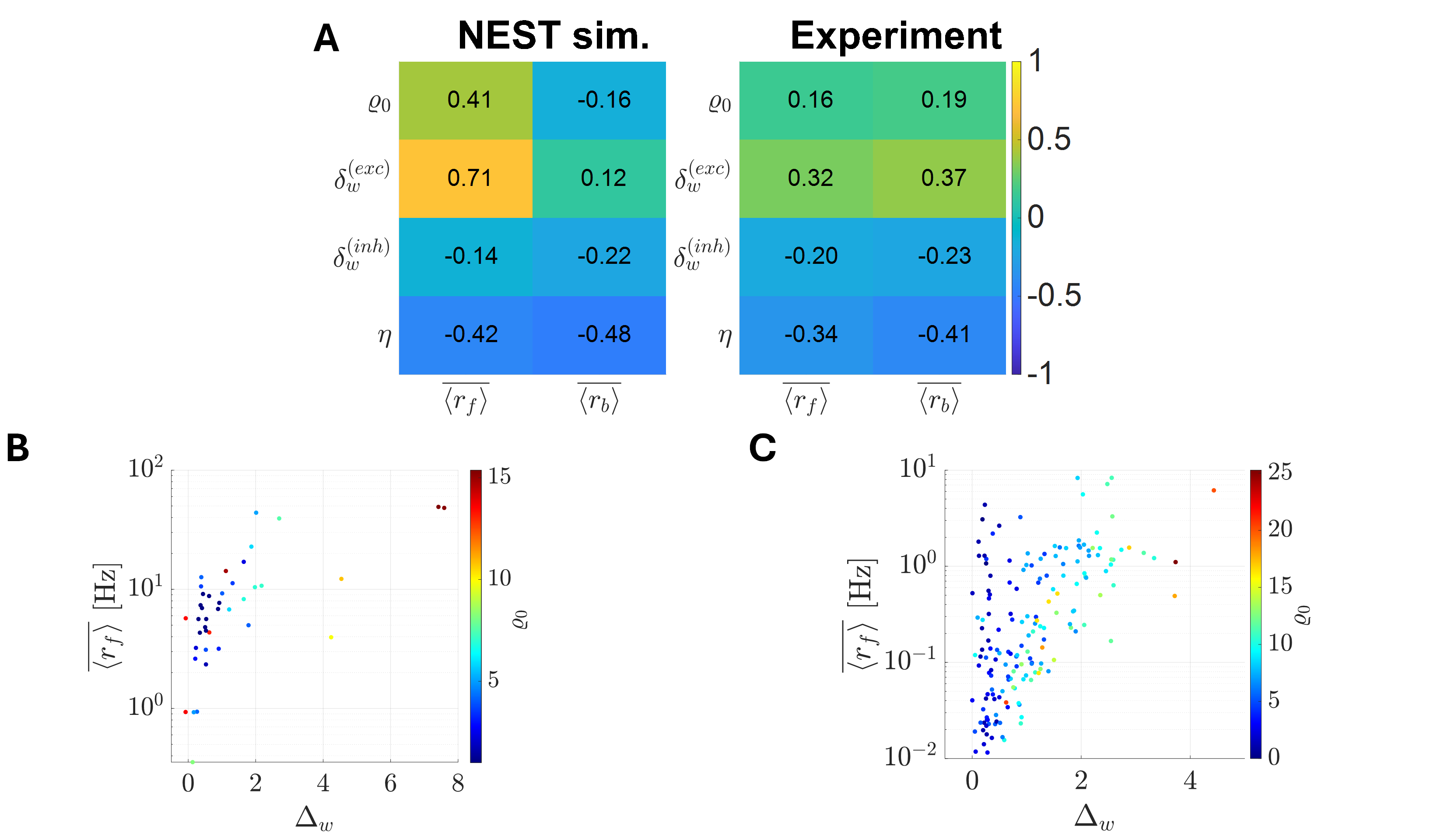}
    \caption{\textbf{Dependence of culture-level observables on global connectivity properties}. (A) Pearson correlation map between ground-truth graph measures (vertical axis) and culture observables (horizontal axis) for both NEST simulations and MEA experiments. (B)–(C) Mean firing rate (MFR, $\overline{\langle r_f \rangle}$) as a function of the effective excitation–inhibition drive, $\Delta_w=\delta_w^{(\mathrm{exc})}-\delta_w^{(\mathrm{inh})}$, for NEST simulations (B) and MEA experiments (C). We observe an exponential increase of the MFR with respect to $\Delta_w$ (note that the figures are shown on a logarithmic scale), which can be well approximated by a gain model of the form $\sim e^{g \Delta_w}$. This amplification is strongly modulated by network complexity, quantified by the relative spectral radius $\varrho_0$ (with individual networks color-coded according to $\varrho_0$). Specifically, networks with lower relative complexity exhibit a larger effective gain, leading to an earlier onset and steeper increase in activity. In contrast, more complex networks display a more gradual and weaker nonlinear response.
    }
    \label{fig:global}
\end{figure}

In this section, we examine how global connectivity properties relate to culture-level observables. Specifically, we investigate how the overall activity intensity is shaped by global graph measures. The relationships between these variables are summarized in Fig.~\ref{fig:global}.

%ADDED
As expected, spike and burst activity correlate positively with the global excitatory weight $\delta_w^{(\mathrm{exc})}$ and negatively with the global inhibitory weight $\delta_w^{(\mathrm{inh})}$ (Fig. \ref{fig:global}A), for both, NEST simulation and experimental data. We hypothesize that this dependence follows an amplification-like nonlinear relationship, in which activity grows approximately exponentially with the effective excitation–inhibition drive, defined as $\Delta_w = \delta_w^{(\mathrm{exc})}-\delta_w^{(\mathrm{inh})}$. Specifically, activity may scale as $\sim e^{g\Delta_w}$, where $g$ represents an effective network gain. We further find that this gain is modulated by network complexity, quantified by the spectral radius $\varrho_0$. A plausible interpretation is that lower-complexity networks operate at higher effective gain, leading to stronger recurrent amplification for a given excitation–inhibition drive, whereas richer connectivity distributes activity over multiple pathways and reduces the net amplification (Fig. \ref{fig:global}B–C).

\subsection{Node Centrality Measures as Predictors of Local Physical Activity}
\begin{figure}
    \centering
    \includegraphics[width=0.98\linewidth]{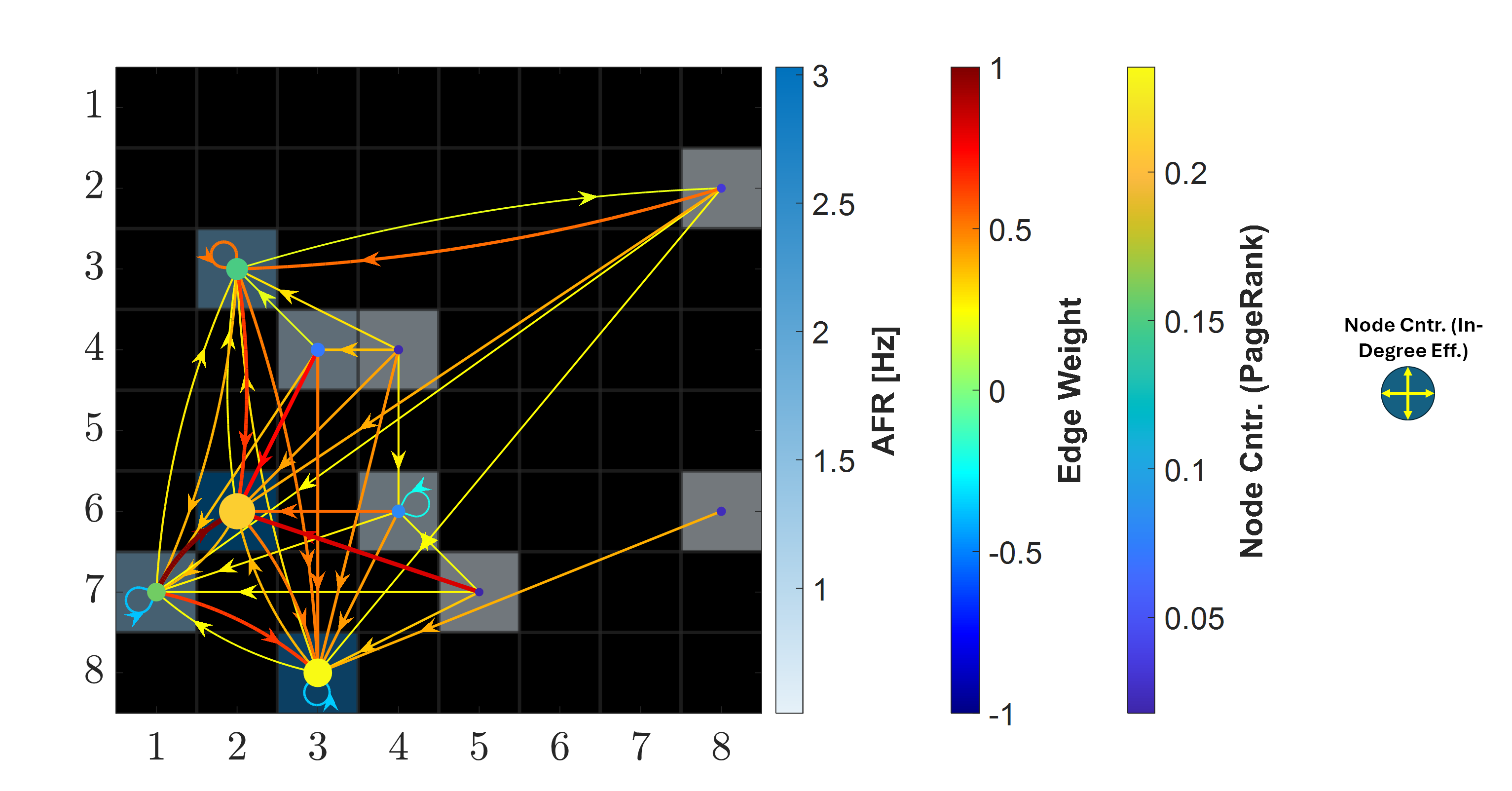}
    \caption{\textbf{Example map of an analyzed neuronal culture obtained from a MEA experiment}. The $8 \times 8$ grid corresponds to a MEA chip with 60 electrodes (64 positions excluding the four corner sites), where each pixel represents an electrode and, correspondingly, a node in the network. The color-coded squares indicate the measured average firing rate (AFR) at each electrode. Superimposed on the grid is the connectivity map inferred by the RC model via the ICM. Nodes are shown as circles and directed edges as arrows, with edge weight encoded by both color (magnitude and sign) and width (magnitude). Node centralities are represented by node size and color. For illustration, two centrality measures are displayed: PageRank (color) and effective in-degree (node diameter).}
    \label{fig:example map}
\end{figure}

\begin{figure}
    \centering
    \includegraphics[width=1\linewidth]{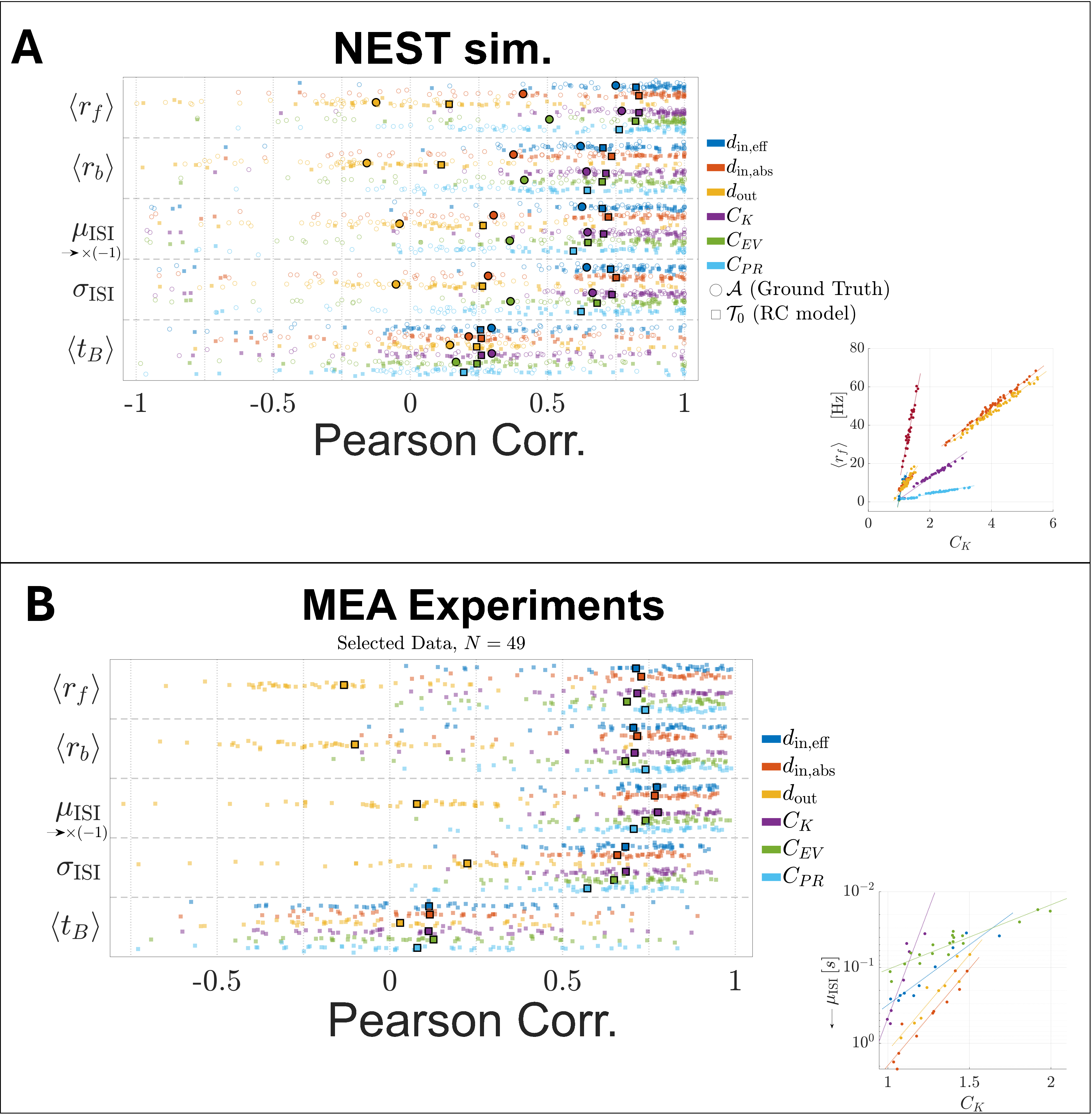}
    \caption{\textbf{Pearson correlations between node centrality measures and node-level observables.} (A) NEST simulation data; (B) MEA experimental dataset—49 selected recordings (out of 170 total) with the highest data-richness parameter $q$, previously shown to correlate with Reservoir Computing (RC) model performance \cite{AUSLENDER} and, in the present study, likewise associated with stronger correlations at higher $q$. The forest plots in (A) and (B) present the Pearson correlation coefficients. The vertical axis lists node-level observables, while each marker represents the correlation obtained with a specific centrality measure (color-coded). The horizontal axis shows the corresponding correlation values (inverted for $\mu_\mathrm{ISI}$). Large colored markers denote the mean across experiments, while smaller shaded markers correspond to individual simulations/recordings. For NEST simulations, centralities are computed from both the ground-truth adjacency matrix (large circles: mean; small open circles: individual simulations) and the ICM (large squares: mean; small colored squares: individual simulations); for MEA experiments, only ICM-based centralities are available. \textbf{Small insets on the buttom right:} Representative scatter plots for a selected centrality–observable pair (best-performing combination). In (A): average firing rate (AFR) vs. Katz centrality, shown for the 10 simulations with the highest correlations. In (B): characteristic ISI timescale in logarithmic scale (y-axis inverted), vs. Katz centrality, shown for the 5 experiments with the highest correlations. In both panels, a linear regression fit is overlaid for each dataset.}
    \label{fig:nodes corrs}
\end{figure}

We evaluated several node centrality measures, namely in- and out-degree, Katz, eigenvector, and PageRank centralities, against node-level activity metrics computed for each node. Across both simulated and experimental datasets, we observe consistently strong associations between centrality and local dynamics. Fig.~\ref{fig:example map} illustrates a representative experimental network, while Fig.~\ref{fig:nodes corrs} summarizes the correlations between centrality measures and node-level observables for both NEST simulations and MEA recordings.

For the experimental data, we further refined the dataset by excluding recordings exhibiting weak correlations. Such cases were predominantly associated with low values of the training-data richness parameter \( q = \frac{\#\text{training time samples}}{N} \), where \(N\) denotes the number of nodes \cite{AUSLENDER}. This parameter has previously been shown to correlate with Reservoir Computing (RC) model performance; accordingly, we attribute the weaker observed correlations primarily to datasets that also yield poorer RC performance. Notably, in this analysis the data points are node-based, resulting in substantially larger sample sizes and ensuring robust statistical support despite the filtering procedure.

For both simulations and experiments, we found that some of the node centralities correlate substantially with the measured observables, with the notable exception of the average burst length, which exhibited considerably weaker correlations (Fig.~\ref{fig:nodes corrs} upper panels). In contrast, spike and burst rates, as well as inter-spike interval (ISI) statistics, show relatively strong correlations with most of the investigated centrality measures. These results suggest that structural properties of the network, as captured by node centralities, are closely related to the local dynamical behavior of the corresponding neuronal populations.

For the NEST simulation results (Fig.~\ref{fig:nodes corrs}A), correlations can be computed using centralities derived either from the ground-truth adjacency matrix $\mathcal{A}$ or from the ICM $\mathcal{T}_0$, inferred by the RC model from the spiking signals. Interestingly, in the vast majority of cases the correlations obtained from the ICM are higher than those computed from the ground-truth connectivity. One possible explanation is that inhibitory connections are somewhat harder to recover by the RC model (Fig.~\ref{fig: weights prediction}; Table~\ref{tab: ICM accuracy results}). As a result, the inferred connectivity may partially redistribute inhibitory pathways through alternative effective connections, which also explains the higher interpretation of the spectral radius by the RC-model (Fig. \ref{fig: weights prediction}D), leading to a representation that is closer to a linearized interaction structure. This effect is particularly evident in the large difference observed in the correlation between the absolute in-degree ($d_{\mathrm{in,abs}}$) and the firing rate $\langle r_f \rangle$ when comparing $\mathcal{A}$ and $\mathcal{T}_0$. In the ground-truth network, inhibitory inputs reduce this correlation for $d_{\mathrm{in,abs}}$, whereas they contribute to the correlation when the effective in-degree ($d_{\mathrm{in,eff}}$) is considered. Overall, the ICM appears to slightly rearrange the connectivity relative to the ground truth, likely due to the linearization inherent in the RC model, in a way that produces centrality measures more related to the observed activity in the first-order interaction.

We note that, in the vast majority of cases, the in-degree exhibits a stronger correlation with node activity than the out-degree. In contrast, the correlations associated with the out-degree are distributed across both positive and negative values, resulting in a less consistent overall relationship. This variability likely reflects the diversity of the network architectures, as outgoing connections can influence the network dynamics in substantially different ways depending on the local circuitry and the functional role of the target nodes. 

Since our analysis focuses on the relationship between a node’s centrality and its own activity, the out-degree is generally expected to have only a limited direct effect on the originating node, except in the presence of self-loops. In the ground-truth adjacency matrices, self-loops were not included. Although intra-population recurrent connections were present in the simulations, they were not represented as self-connections in the adjacency matrices. In contrast, the reservoir computing (RC) model is capable of inferring self-loops. Consistently, the results shown in Fig.~\ref{fig:nodes corrs}A indicate that the correlations obtained from the RC-inferred connectivity are shifted toward more positive values relative to those derived from the ground-truth connectivity. Nevertheless, strong negative correlations may also emerge, as observed both in the simulations, particularly for the ground-truth networks, and in the experimental data, potentially due to short-range inhibitory interactions.

Another notable observation is that, in the experimental data, the mean inter-spike interval (ISI) exhibits stronger correlations with the centrality measures than the firing rate, in contrast to the simulations. This suggests that, in MEA recordings, where the measured activity reflects aggregated population-level dynamics rather than single-neuron activity, ISI-based statistics may provide a more informative characterization of the local dynamics than the average firing rate alone, which constitutes a comparatively coarse descriptor.

\section{Discussion}

This work provides an in-depth analysis of neuronal cultures subjected to multichannel electrophysiological recordings, moving beyond conventional metrics that characterize only local activity and lack a direct interpretation at the network level. The analysis is based on a recently reported RC model which, in addition to modeling the macroscopic dynamics of the cultured network, enables the inference of its underlying connectivity map with high accuracy under specific conditions \cite{AUSLENDER}. This connectivity map, formalized as ICM, was subsequently treated as the network adjacency matrix and used to compute graph-theoretical measures, enabling a quantitative interpretation of node centrality and network organization. In the following, we discuss the main findings of this work and examine their implications.

\subsection{Structural and Dynamical Factors Governing ICM–Ground-Truth Agreement}
To provide a foundation for the analyses that follow, we first investigated how strongly the ICM correlates with the ground-truth adjacency matrix as a function of the underlying graph-theoretical features. To this end, we employed NEST simulations described in detail in \cite{AUSLENDER}, in which the network architecture is defined by fixed weighted inhibitory and excitatory connections. The ICM, in contrast, is obtained by training the RC model on the emergent spatio-temporal spike trains, followed by linearization of the trained model. In essence, the ICM characterizes \textit{Effective Connectivity} rather than \textit{Functional Connectivity}, as it is specifically designed to infer causal relationships between nodes from short-time-window signal sequences by learning the underlying dynamical interactions. Nevertheless, a direct and universal correspondence between the ground-truth adjacency matrix and the ICM cannot be assumed. First, the degree of correlation depends on the quality of the model training, which requires appropriate adaptation of the full nonlinear RC architecture through its internal hyperparameters to the observed dynamics, a task that can be nontrivial in practice. Second, the linearization procedure reveals connectivity only at the level of first-order interactions, isolating both the nonlinear reservoir function and synaptic plasticity effects that are themselves influenced by higher-order network pathways. As a result, the RC model may encode interaction pathways in a manner that deviates from the purely structural organization of the underlying network. Hence, a low correlation between the ICM and the ground-truth matrix does not necessarily imply poor performance of the RC model. Model performance is more appropriately assessed through the training and validation losses of the full model, as well as the connectivity confidence (Eq.~\ref{eq: conn. conf.}) of the linearized model. Rather, a low correlation indicates that the network dynamics are sufficiently intricate, and dominated by nonlinear relations, that linearization fails to isolate the implicit connections between nodes.

This analysis of RC model performance is therefore complementary to the investigation presented in \cite{AUSLENDER}, where performance was characterized as a function of the number of network nodes (or populations) and the RC model hyperparameters. In the present study, we instead focus on how graph-theoretical measures of the underlying network relate to the observed agreement between the ICM and the ground-truth connectivity.

\subsection{An Optimal Balance of Inhibitory Connectivity Is Required for Accurate Connectivity Inference by the ICM}

We find that the inhibitory–excitatory ratio strongly affects the interpretability of the ICM relative to the ground-truth adjacency matrix. This behavior can be understood from the way the ICM is derived, namely through a linearization of the full RC model, in which short-time nonlinear dynamical effects and higher-order interactions are neglected. Networks dominated by excitation are expected to produce strong transient nonlinear responses, enhanced nonlinear effects, particularly near unstable or critical dynamical regimes \cite{hennequin2014optimal,ganguli2008memory}. Accordingly, we observe that for $\eta \approx 0$ the accuracy metrics of the ICM with respect to the ground-truth connectivity tend to be reduced. Conversely, strongly inhibitory networks can also introduce nonlinear dynamical effects and nonlinear response properties \cite{tsodyks1997paradoxical}, and in this regime the reconstruction accuracy similarly declines. 

These observations suggest the existence of an optimal inhibitory level at which the network dynamics remain sufficiently close to linear within a finite time window, allowing the linearization step to correctly infer the effective interactions between nodes. This interpretation is consistent with previous studies showing that balanced excitation and inhibition can stabilize network activity and support tractable low-dimensional or weakly nonlinear dynamics \cite{vanvreeswijk1996chaos,renart2010asynchronous}. This interpretation holds provided that the full nonlinear RC model itself is well trained and reproduces the observed activity with adequate accuracy. 

Evidence for this behavior can be observed both at the aggregate level, across all weights of the tested networks, in Table~\ref{tab: ICM accuracy results} and Figs.~\ref{fig: weights prediction}A–C, and at the level of individual networks, as illustrated in Figs.~\ref{fig: weights prediction}E and \ref{fig:performance}A, C (blue line).

\subsection{RC Model Performance Declines for Highly Complex Networks}

Our analysis shows that network complexity, quantified by the spectral radius $\varrho_0$ and the degree density $\delta_w$, directly affects RC-model performance in terms of training, validation, and stability. This behavior can be intuitively understood: the RC model is trained to capture causal interaction patterns in the network dynamics; increased circuit complexity is generally associated with a broader repertoire of spatiotemporal dynamics \cite{voges2012complex,sussillo2009generating}, making the effective interaction pathways more difficult to infer. This effect is illustrated in Figs.~\ref{fig:performance}A, B, and D. By contrast, the inhibitory level $\eta$ appears to have a weaker influence on the stability and overall performance of the RC model (Figs.~\ref{fig:performance}A, C).

Naturally, when dealing with experimental data, the performance of the RC model strongly depends on data quality. Experimental recordings may be affected by various sources of noise, arising, for example, from limitations of the experimental setup or from inadequate adaptation of the biological sample to the recording conditions. Such effects can lead to over- or under-detection of spikes, thereby distorting the temporal information required for accurate network inference. Consequently, the reliability of the present analysis depends critically on the entire preprocessing and acquisition pipeline preceding the inference stage.

\subsection{Graph Properties Modulate the Overall Activity of the Culture}
Overall, as shown in Fig.~\ref{fig:global}, the synaptic weight structure appears to play a dominant role in determining network excitability, with inhibitory coupling acting to constrain population activity. This observation should be regarded primarily as qualitative rather than quantitative. From the perspective of the ground-truth adjacency matrices, the result is expected, as the synaptic weights that govern the dynamics naturally determine the global activity patterns. However, from the perspective of the RC-inferred connectivity, and particularly in the case of experimental data, the result is considerably more informative. Specifically, the ability of the RC model to infer excitatory and inhibitory pathways with relatively high accuracy provides additional validation that the inferred connectivity captures functionally relevant aspects of the underlying network dynamics.

\subsection{Node Centrality Measures Show an Overall Association with Local Physical Observables}
In this section, we demonstrated that several node centrality measures correlate with locally measured electrophysiological observables, particularly neuronal firing activity. A related analysis was previously reported in Ref.~\cite{Fletcher}, where spike rate was correlated with different centrality measures in purely \textit{in silico} neuronal networks. The authors found that, in the general case, Katz centrality exhibited the strongest correlation with firing rate and, notably, remained the only centrality measure preserving a significant correlation in networks with substantial inhibitory connectivity.

In contrast, our results indicate that additional centrality measures, including in-degree, PageRank, and eigenvector centrality, also exhibit substantial correlations with firing rate, even in networks containing non-negligible inhibitory interactions. Several methodological differences between the two studies may explain this discrepancy.

The first major difference concerns the network representation itself. In Ref.~\cite{Fletcher}, each node represents a single neuron that is either exclusively excitatory or exclusively inhibitory. In our framework, however, the network is represented under MEA measurement conditions, such that each node corresponds to a small neuronal population rather than an individual neuron. Consequently, each node effectively receives both excitatory and inhibitory inputs. Empirically, most nodes in our constructed and inferred networks remain predominantly excitatory, which may explain the partial agreement between our findings and those reported in Ref.~\cite{Fletcher} for excitation-dominated networks.

A second important difference arises from the neuronal models used in the simulations. Ref.~\cite{Fletcher} employed the Generalized Leaky Integrate-and-Fire (GLIF) model, whereas our simulations are based on the Izhikevich neuron model implemented within the NEST framework \cite{izhikevich2004model}. The GLIF model extends the classical leaky integrate-and-fire formalism by incorporating additional biologically motivated mechanisms, such as adaptive thresholds and after-spike dynamics, while maintaining relatively low computational complexity. In contrast, the Izhikevich model describes neuronal activity through a coupled nonlinear dynamical system capable of reproducing a broad range of biologically realistic firing patterns with high computational efficiency. These differences in the underlying neuronal dynamics may contribute to the distinct correlation structures observed in the two studies, particularly in the context of population-level activity and recurrent network interactions.

An additional distinction concerns the stimulation protocol used to drive network activity. In Ref.~\cite{Fletcher}, neuronal activity was generated by externally applied fixed stimuli delivered uniformly across the network. In contrast, our simulations were designed to reproduce spontaneous activity regimes. To this end, each neuron was driven by Gaussian noise representing membrane potential fluctuations, together with Poisson-distributed spike trains applied either globally or to individual neurons within a population. This approach produces a more heterogeneous and distributed driving mechanism, under which neuronal activity may more directly reflect the underlying network topology encoded by the different centrality measures.

Nevertheless, although Katz centrality does not exhibit a clear superiority in terms of average correlation values in our analysis, its correlation distribution appears more densely concentrated at high positive values (e.g., $>0.75$) relative to the other centrality measures. This observation may indicate partial agreement with the conclusions of Ref.~\cite{Fletcher}.

For comparison, we also evaluated the same correlations using the full simulated networks rather than the simplified MEA-based representation, such that all neurons within each population were represented individually together with their corresponding firing rates. The results, shown in Table \ref{tab:full explicit network results}, reveal systematically lower correlations compared to the MEA representation, although the correlations remain sufficiently strong to preserve statistical significance. This reduction may partly originate from the stochastic noise introduced into the simulations. More importantly, it suggests that the effective population-level clustering inherent to the MEA representation produces nodes that are more representative of the collective local dynamics.

Overall, despite the methodological differences between the two studies, both works support the existence of a meaningful relationship between structural network properties and local neuronal activity. While the precise hierarchy of the centrality measures may depend on the neuronal model, network representation, and stimulation protocol, the consistent emergence of strong correlations across different conditions suggests that centrality-based descriptors capture functionally relevant aspects of the underlying dynamics.

Nevertheless, the principal contribution of the present study lies in extending these analyses to experimental data through a validated connectivity inference framework.

\begin{table}[h]
    \caption{Pearson correlations between firing rate and node centrality measures computed for the full simulated NEST networks, where each node represents an individual point-process neuron modeled using the Izhikevich formalism \cite{izhikevich2004model}. Unlike the main results presented in this work, which are based on an MEA-inspired population-level representation, the results shown here correspond to the explicit neuron-level network representation.}
    \centering
    \begin{tabular}{|c c c c c c|}
        \hline
        \multicolumn{6}{|c|}{\makecell{\textbf{Pearson correlation between AFR and node centrality} \\ (mean $\pm$ std)}} \\
        \hline
         $d_{\mathrm{in,eff}}$ & $d_{\mathrm{in,abs}}$ & $d_{\mathrm{out}}$ & $C_K$ & $C_{EV}$ & $C_{PR}$  \\
         $0.258 \pm0.217$ & $0.105 \pm 0.324$ & $-0.033 \pm 0.231$ & $0.252 \pm 0.257$ & $0.190 \pm 0.335$ & $-0.042 \pm 0.230$\\
         \hline         
    \end{tabular}
    
    \label{tab:full explicit network results}
\end{table}

\subsection{Further Evidence Supporting RC Model’s Validity}
In this study, we demonstrate how well-established theoretical principles can be applied directly to experimental neuronal activity data through the recently developed RC framework, which enables data-driven reconstruction of network connectivity. The framework thereby gains additional validation: the connectivity map obtained through the ICM is not only benchmarked against known ground-truth adjacency matrices in simulations, but is also shown to be consistent with experimentally measured observables in real neuronal networks. Although some of the correlations identified here may appear mathematically expected from an analytical perspective, the RC model remains valuable in that it correctly recovers and interprets these network relationships directly from activity data, without prior access to the underlying connectivity. This supports the RC approach as a meaningful tool for network inference beyond merely reproducing trivial structural dependencies.

\subsection{ICM as an Adjacency Matrix}

The simulation results indicate that the ICM representation generally provides a meaningful description of network properties, as reflected by graph measures that successfully interpret aspects of the network dynamics. As shown above, the ICM does not always reproduce the exact structural connectivity (e.g., low $\rho_C$, Fig.~\ref{fig:performance}C). In some cases it may misestimate the inhibition level (Fig.~\ref{fig: weights prediction}D) or the local circuitry complexity (Fig.~\ref{fig: weights prediction}E). Nevertheless, the graph-theoretic properties derived from the ICM exhibit consistent and interpretable behavior throughout the study.

\subsection{Simulations Vs. Experiments}
Overall, the experimental results are largely consistent with those obtained from simulations, with a few deviations that likely arise from more complex or heterogeneous network topologies present in the biological data. Notably, the correlations obtained from experimental recordings are slightly lower than those observed in the \textit{in-silico} networks. This difference is expected, as each electrode samples only a partial subset of the surrounding neuronal population and therefore does not capture the full connectivity structure \cite{Obien2015}. In addition, signal mixing may occur when the local cell density is high, leading to partially entangled recordings. From this perspective, improved results could likely be achieved using higher-density recording systems, such as HD-MEAs \cite{emmenegger2019technologies}. 

It should also be noted that the effectiveness of this approach depends critically on the training procedure of the RC model, including the selection of hyperparameters and architectural settings. Consequently, questions regarding the optimal adaptation of the RC framework to different datasets and dynamical regimes remain open for future investigation.

\subsection{Implications for Network-Level Analysis of Neuronal Cultures and Concluding Remarks}

This study provides a framework for analyzing neuronal cultures beyond purely local electrophysiological measurements by incorporating network-level descriptors derived from graph theory. By reconstructing an effective connectivity map through the RC model and interpreting it through graph-theoretical metrics, the proposed approach enables a more comprehensive characterization of the functional organization of neuronal cultures. In particular, it allows one to relate structural properties of the network to the observed dynamics at both the node level and the culture level.

Such an approach may be valuable in several neurobiological contexts. In particular, it could be applied in neuropharmacological studies, where pharmacological interventions often alter synaptic transmission and circuit-level interactions rather than only modifying local firing statistics. In these cases, changes in neuronal pathways and connectivity patterns may manifest not only in local activity measures but also in the global topology and organization of the inferred network. The ability to quantify connectivity-related properties, such as centrality, connectivity balance, and network complexity, may therefore provide additional indicators of functional changes induced by experimental manipulations.

More broadly, the present framework demonstrates how connectivity inference combined with graph-theoretical analysis can provide interpretable network-level descriptions of neuronal dynamics in \textit{in vitro} systems. In principle, the same conceptual methodology could also be extended to \textit{in vivo} recordings obtained through large-scale electrophysiological or imaging techniques, where the relationship between network structure and activity remains a central open question. From this perspective, the RC-based approach may offer a promising tool for bridging experimentally measured neural activity with the underlying network organization across different biological scales.

\section*{Acknowledgments}
This work was financed by the European Union - NextGenerationEU and National Recovery and Resilience Plan (NRRP) - Mission 4 Component 2 Investment 1.2 – “Funding projects presented by young researchers” MSCA PNRR Young Researchers, “CIRCUS project” - MSCA20240000106 - CUP E63C25000820007.
\printbibliography

@article{AUSLENDER,
title = {Decoding neuronal networks: A Reservoir Computing approach for predicting connectivity and functionality},
journal = {Neural Networks},
volume = {184},
pages = {107058},
year = {2025},
issn = {0893-6080},
doi = {https://doi.org/10.1016/j.neunet.2024.107058},
url = {https://www.sciencedirect.com/science/article/pii/S0893608024009870},
author = {Ilya Auslender and Giorgio Letti and Yasaman Heydari and Clara Zaccaria and Lorenzo Pavesi}
}

@article{Fletcher,
author = {Fletcher, Jack McKay and Wennekers, Thomas},
title = {From Structure to Activity: Using Centrality Measures to Predict Neuronal Activity},
journal = {International Journal of Neural Systems},
volume = {28},
number = {02},
pages = {1750013},
year = {2018},
doi = {10.1142/S0129065717500137},
    note ={PMID: 28076982},

URL = { 
    
        https://doi.org/10.1142/S0129065717500137
    
    

},
eprint = { 
    
        https://doi.org/10.1142/S0129065717500137
    
    

}
}

@article{katz1953new,
  title={A new status index derived from sociometric analysis},
  author={Katz, Leo},
  journal={Psychometrika},
  volume={18},
  number={1},
  pages={39--43},
  year={1953},
  publisher={Springer-Verlag}
}

@article{Nelson2021,
  title={Neuronal Graphs: A Graph Theory Primer for Microscopic, Functional Networks},
  author={Nelson, Carl J. and Bonner, Stephen},
  journal={Frontiers in Neural Circuits},
  year={2021},
  doi={10.3389/fncir.2021.662882}
}

@article{Schroeter2017,
  title={Micro-connectomics: probing the organization of neuronal networks at the cellular scale},
  author={Schr{\"o}ter, Manuel and Paulsen, Ole and Bullmore, Edward T.},
  journal={Nature Reviews Neuroscience},
  year={2017},
  volume={18},
  pages={131-146},
  doi={10.1038/nrn.2016.182}
}

@article{sporns2012simple,
  title={From simple graphs to the connectome: networks in neuroimaging},
  author={Sporns, Olaf},
  journal={Neuroimage},
  volume={62},
  number={2},
  pages={881--886},
  year={2012},
  publisher={Elsevier},
  doi = {10.1016/j.neuroimage.2011.08.085}
}

@article{de2018connectivity,
  title={Connectivity inference from neural recording data: Challenges, mathematical bases and research directions},
  author={de Abril, Ildefons Magrans and Yoshimoto, Junichiro and Doya, Kenji},
  journal={Neural Networks},
  volume={102},
  pages={120--137},
  year={2018},
  publisher={Elsevier},
  doi = {10.1016/j.neunet.2018.02.016}
}

@article{Bullmore2012,
  title={The economy of brain network organization},
  author={Bullmore, Ed and Sporns, Olaf},
  journal={Nature Reviews Neuroscience},
  volume={13},
  number={5},
  pages={336--349},
  year={2012},
  doi={10.1038/nrn3214}
}

@article{Sporns2014,
  title={Contributions and challenges for network models in cognitive neuroscience},
  author={Sporns, Olaf},
  journal={Nature Neuroscience},
  volume={17},
  number={5},
  pages={652--660},
  year={2014},
  doi={10.1038/nn.3690}
}

@article{Smith2018,
  title={Distributed network interactions and their emergence in developing neocortex},
  author={Smith, Gordon B. and Hein, Bettina and Whitney, David E. and Fitzpatrick, David and Kaschube, Matthias},
  journal={Nature Neuroscience},
  volume={21},
  number={11},
  pages={1600--1608},
  year={2018},
  doi={10.1038/s41593-018-0247-5}
}

@article{Paninski2018,
  title={Neural data science: accelerating the experiment-analysis-theory cycle in large-scale neuroscience},
  author={Paninski, Liam and Cunningham, John P.},
  journal={Current Opinion in Neurobiology},
  volume={50},
  pages={232--241},
  year={2018},
  doi={10.1016/j.conb.2018.04.007}
}

@article{Spira2013,
  title={Multi-electrode array technologies for neuroscience and cardiology},
  author={Spira, Micha E. and Hai, Amit},
  journal={Nature Nanotechnology},
  volume={8},
  number={2},
  pages={83--94},
  year={2013},
  doi={10.1038/nnano.2012.265}
}

@article{Buzsaki2004LargeScale,
  title   = {Large-scale recording of neuronal ensembles},
  author  = {Buzs{\'a}ki, Gy{\"o}rgy},
  journal = {Nature Neuroscience},
  volume  = {7},
  number  = {5},
  pages   = {446--451},
  year    = {2004},
  doi     = {10.1038/nn1233}
}

@article{Obien2015,
  title={Revealing neuronal function through microelectrode array recordings},
  author={Obien, M.E.J. and Deligkaris, K. and Bullmann, T. and Bakkum, D.J. and Frey, U.},
  journal={Frontiers in Neuroscience},
  volume={8},
  pages={423},
  year={2015},
  doi={10.3389/fnins.2014.00423}
}

@article{Maccione2012,
  title={Multiscale functional connectivity estimation on low-density neuronal cultures recorded by high-density CMOS microelectrode arrays},
  author={Maccione, Alessandro and Garofalo, Matteo and Nieus, Thierry and Tedesco, Mariateresa and Berdondini, Luca and Martinoia, Sergio},
  journal={Journal of Neuroscience Methods},
  volume={207},
  number={2},
  pages={161--171},
  year={2012},
  doi={10.1016/j.jneumeth.2012.04.002}
}

@article{Chiappalone2008,
  title={Network plasticity in cortical assemblies},
  author={Chiappalone, Michela and Massobrio, Paolo and Martinoia, Sergio},
  journal={European Journal of Neuroscience},
  volume={28},
  pages={221--237},
  year={2008},
  doi={10.1111/j.1460-9568.2008.06259.x}
}

@article{Vicente2011,
  title={Transfer entropy—a model‑free measure of effective connectivity for the neurosciences},
  author={Vicente, R. and Wibral, M. and Lindner, M. and Pipa, G.},
  journal={Journal of Computational Neuroscience},
  volume={30},
  pages={45--67},
  year={2011},
  doi={10.1007/s10827-010-0262-3}
}

@article{Stetter2012,
  title={Model-free reconstruction of excitatory neuronal connectivity from calcium imaging signals},
  author={Stetter, O. and Battaglia, D. and Soriano, J. and Geisel, T.},
  journal={PLoS Computational Biology},
  volume={8},
  number={8},
  pages={e1002653},
  year={2012},
  doi={10.1371/journal.pcbi.1002653}
}

@article{Poli2015,
  title={Functional connectivity in in vitro neuronal assemblies},
  author={Poli, Daniele and Pastore, Vito P. and Massobrio, Paolo},
  journal={Frontiers in Neural Circuits},
  volume={9},
  pages={57},
  year={2015},
  doi={10.3389/fncir.2015.00057}
}

@article{endo2021convolutional,
  title={A convolutional neural network for estimating synaptic connectivity from spike trains},
  author={Endo, Daisuke and Kobayashi, Ryota and Bartolo, Ramon and Averbeck, Bruno B and Sugase-Miyamoto, Yasuko and Hayashi, Kazuko and Kawano, Kenji and Richmond, Barry J and Shinomoto, Shigeru},
  journal={Scientific Reports},
  volume={11},
  number={1},
  pages={12087},
  year={2021},
  publisher={Nature Publishing Group UK London},
  doi={10.1038/s41598-021-91244-w}
}

@article{Wagenaar2006,
  title={Extremely rich repertoire of bursting patterns during the development of cortical cultures},
  author={Wagenaar, Daniel A. and Pine, Jonathan and Potter, Steve M.},
  journal={BMC Neuroscience},
  volume={7},
  pages={11},
  year={2006},
  doi={10.1186/1471-2202-7-11}
}

@article{pasquale2010self,
  title={A self-adapting approach for the detection of bursts and network bursts in neuronal cultures},
  author={Pasquale, Valentina and Martinoia, Sergio and Chiappalone, Michela},
  journal={Journal of computational neuroscience},
  volume={29},
  number={1},
  pages={213--229},
  year={2010},
  publisher={Springer}
}

@article{chiappalone2005burst,
  title={Burst detection algorithms for the analysis of spatio-temporal patterns in cortical networks of neurons},
  author={Chiappalone, Michela and Novellino, Antonio and Vajda, Ildiko and Vato, Alessandro and Martinoia, Sergio and van Pelt, Jaap},
  journal={Neurocomputing},
  volume={65},
  pages={653--662},
  year={2005},
  publisher={Elsevier}
}

@article{antonello2022self,
  title={Self-organization of in vitro neuronal assemblies drives to complex network topology},
  author={Antonello, Priscila C and Varley, Thomas F and Beggs, John and Porcionatto, Marim{\'e}lia and Sporns, Olaf and Faber, Jean},
  journal={Elife},
  volume={11},
  pages={e74921},
  year={2022},
  publisher={eLife Sciences Publications Limited}
}

@article{izhikevich2003simple,
  title={Simple model of spiking neurons},
  author={Izhikevich, Eugene M},
  journal={IEEE Transactions on neural networks},
  volume={14},
  number={6},
  pages={1569--1572},
  year={2003},
  publisher={IEEE}
}

@article{izhikevich2004model,
  title={Which model to use for cortical spiking neurons?},
  author={Izhikevich, Eugene M},
  journal={IEEE transactions on neural networks},
  volume={15},
  number={5},
  pages={1063--1070},
  year={2004},
  publisher={Ieee}
}

@article{Page1998,
  title={The anatomy of a large-scale hypertextual Web search engine},
  author={Brin, Sergey and Page, Lawrence},
  journal={Computer Networks and ISDN Systems},
  volume={30},
  number={1--7},
  pages={107--117},
  year={1998},
  doi={10.1016/S0169-7552(98)00110-X}
}

@article{Bonacich1972,
  title={Factoring and weighting approaches to status scores and clique identification},
  author={Bonacich, Phillip},
  journal={Journal of Mathematical Sociology},
  volume={2},
  number={1},
  pages={113--120},
  year={1972},
  doi={10.1080/0022250X.1972.9989806}
}

@article{Bonacich1987,
  title={Power and centrality: A family of measures},
  author={Bonacich, Phillip},
  journal={American Journal of Sociology},
  volume={92},
  number={5},
  pages={1170--1182},
  year={1987},
  doi={10.1086/228631}
}

@article{Swets1988,
  title={Measuring the accuracy of diagnostic systems},
  author={Swets, John A.},
  journal={Science},
  volume={240},
  number={4857},
  pages={1285--1293},
  year={1988},
  doi={10.1126/science.3287615}
}

@inproceedings{Davis2006,
  title={The relationship between Precision-Recall and ROC curves},
  author={Davis, Jesse and Goadrich, Mark},
  booktitle={Proceedings of the 23rd International Conference on Machine Learning},
  pages={233--240},
  year={2006},
  doi={10.1145/1143844.1143874}
}

@book{Rijsbergen1979,
  title={Information Retrieval},
  author={van Rijsbergen, Cornelis J.},
  publisher={Butterworths},
  year={1979}
}

@article{Powers2011,
  title={Evaluation: From precision, recall and F-measure to ROC, informedness, markedness and correlation},
  author={Powers, David M. W.},
  journal={Journal of Machine Learning Technologies},
  volume={2},
  number={1},
  pages={37--63},
  year={2011}
}

@book{Kandel2013Principles,
  title     = {Principles of Neural Science},
  author    = {Kandel, Eric R. and Schwartz, James H. and Jessell, Thomas M. and Siegelbaum, Steven A. and Hudspeth, A. J.},
  edition   = {5},
  year      = {2013},
  publisher = {McGraw-Hill},
  address   = {New York}
}

@article{BassettSporns2017NetworkNeuroscience,
  title   = {Network neuroscience},
  author  = {Bassett, Danielle S. and Sporns, Olaf},
  journal = {Nature Neuroscience},
  volume  = {20},
  number  = {3},
  pages   = {353--364},
  year    = {2017},
  doi     = {10.1038/nn.4502}
}

@article{BullmoreSporns2009Complex,
  title   = {Complex brain networks: graph theoretical analysis of structural and functional systems},
  author  = {Bullmore, Ed and Sporns, Olaf},
  journal = {Nature Reviews Neuroscience},
  volume  = {10},
  number  = {3},
  pages   = {186--198},
  year    = {2009},
  doi     = {10.1038/nrn2575}
}

@article{RubinovSporns2010ComplexMeasures,
  title   = {Complex network measures of brain connectivity: uses and interpretations},
  author  = {Rubinov, Mikail and Sporns, Olaf},
  journal = {NeuroImage},
  volume  = {52},
  number  = {3},
  pages   = {1059--1069},
  year    = {2010},
  doi     = {10.1016/j.neuroimage.2009.10.003}
}

@article{Crossley2014HubsDisorders,
  title   = {The hubs of the human connectome are generally implicated in the anatomy of brain disorders},
  author  = {Crossley, Nicholas A. and Mechelli, Andrea and Scott, James and Carletti, Francesca and Fox, Peter T. and McGuire, Philip and Bullmore, Edward T.},
  journal = {Brain},
  volume  = {137},
  number  = {8},
  pages   = {2382--2395},
  year    = {2014},
  doi     = {10.1093/brain/awu132}
}

@article{Sporns2005Connectome,
  title   = {The human connectome: A structural description of the human brain},
  author  = {Sporns, Olaf and Tononi, Giulio and K{\"o}tter, Rolf},
  journal = {PLoS Computational Biology},
  volume  = {1},
  number  = {4},
  pages   = {e42},
  year    = {2005},
  doi     = {10.1371/journal.pcbi.0010042}
}

@article{VanEssen2013HCP,
  title   = {The WU-Minn Human Connectome Project: An overview},
  author  = {Van Essen, David C. and Smith, Stephen M. and Barch, Deanna M. and Behrens, Timothy E. J. and Yacoub, Essa and Ugurbil, Kamil},
  journal = {NeuroImage},
  volume  = {80},
  pages   = {62--79},
  year    = {2013},
  doi     = {10.1016/j.neuroimage.2013.05.041}
}

@article{Fornito2015Disorders,
  title   = {The connectomics of brain disorders},
  author  = {Fornito, Alex and Zalesky, Andrew and Breakspear, Michael},
  journal = {Nature Reviews Neuroscience},
  volume  = {16},
  number  = {3},
  pages   = {159--172},
  year    = {2015},
  doi     = {10.1038/nrn3901}
}

@book{Diestel2017GraphTheory,
  title     = {Graph Theory},
  author    = {Diestel, Reinhard},
  edition   = {5},
  year      = {2017},
  publisher = {Springer},
  address   = {Berlin},
  doi       = {10.1007/978-3-662-53622-3}
}

@book{Newman2010Networks,
  title     = {Networks: An Introduction},
  author    = {Newman, Mark},
  year      = {2010},
  publisher = {Oxford University Press},
  address   = {Oxford}
}

@article{Logothetis2008FMRI,
  title   = {What we can do and what we cannot do with fMRI},
  author  = {Logothetis, Nikos K.},
  journal = {Nature},
  volume  = {453},
  number  = {7197},
  pages   = {869--878},
  year    = {2008},
  doi     = {10.1038/nature06976}
}

@article{Lichtman2011Connectomics,
  title   = {The big and the small: challenges of imaging the brain's circuits},
  author  = {Lichtman, Jeff W. and Denk, Winfried},
  journal = {Science},
  volume  = {334},
  number  = {6056},
  pages   = {618--623},
  year    = {2011},
  doi     = {10.1126/science.1209168}
}

@article{Pearson1895Regression,
  title   = {Notes on regression and inheritance in the case of two parents},
  author  = {Pearson, Karl},
  journal = {Proceedings of the Royal Society of London},
  volume  = {58},
  pages   = {240--242},
  year    = {1895}
}

@article{hennequin2014optimal,
  title={Optimal control of transient dynamics in balanced networks supports generation of complex movements},
  author={Hennequin, Guillaume and Vogels, Tim P and Gerstner, Wulfram},
  journal={Neuron},
  volume={82},
  number={6},
  pages={1394--1406},
  year={2014},
  publisher={Elsevier},
  doi={10.1016/j.neuron.2014.04.045}
}

@article{ganguli2008memory,
  title={Memory traces in dynamical systems},
  author={Ganguli, Surya and Huh, Dongsung and Sompolinsky, Haim},
  journal={Proceedings of the National Academy of Sciences},
  volume={105},
  number={48},
  pages={18970--18975},
  year={2008},
  publisher={National Academy of Sciences},
  doi={10.1073/pnas.0804451105}
}

@article{vanvreeswijk1996chaos,
  title={Chaos in neuronal networks with balanced excitatory and inhibitory activity},
  author={van Vreeswijk, Carl and Sompolinsky, Haim},
  journal={Science},
  volume={274},
  number={5293},
  pages={1724--1726},
  year={1996},
  publisher={AAAS},
  doi={10.1126/science.274.5293.1724}
}

@article{renart2010asynchronous,
  title={The asynchronous state in cortical circuits},
  author={Renart, Alfonso and de la Rocha, Jaime and Bartho, Peter and Hollender, Liad and Parga, Nestor and Reyes, Alex and Harris, Kenneth D},
  journal={Science},
  volume={327},
  number={5965},
  pages={587--590},
  year={2010},
  publisher={AAAS},
  doi={10.1126/science.1179850}
}

@article{tsodyks1997paradoxical,
  title={Paradoxical effects of external modulation of inhibitory interneurons},
  author={Tsodyks, Misha V and Skaggs, William E and Sejnowski, Terrence J and McNaughton, Bruce L},
  journal={Journal of Neuroscience},
  volume={17},
  number={11},
  pages={4382--4388},
  year={1997},
  publisher={Society for Neuroscience},
  doi={10.1523/JNEUROSCI.17-11-04382.1997}
}

@article{sussillo2009generating,
  title={Generating coherent patterns of activity from chaotic neural networks},
  author={Sussillo, David and Abbott, L. F.},
  journal={Neuron},
  volume={63},
  number={4},
  pages={544--557},
  year={2009},
  publisher={Elsevier},
  doi={10.1016/j.neuron.2009.07.018}
}

@article{voges2012complex,
  title={Complex dynamics in recurrent cortical networks based on spatially realistic connectivities},
  author={Voges, Nicole and Perrinet, Laurent},
  journal={Frontiers in Computational Neuroscience},
  volume={6},
  pages={41},
  year={2012},
  publisher={Frontiers Media},
  doi={10.3389/fncom.2012.00041}
}

@ARTICLE{Gewaltig:NEST,
  author  = {Marc-Oliver Gewaltig and Markus Diesmann},
  title   = {NEST (NEural Simulation Tool)},
  journal = {Scholarpedia},
  year    = {2007},
  volume  = {2},
  pages   = {1430},
  number  = {4}
}

@article{emmenegger2019technologies,
  title={Technologies to study action potential propagation with a focus on HD-MEAs},
  author={Emmenegger, Vishalini and Obien, Marie Engelene J and Franke, Felix and Hierlemann, Andreas},
  journal={Frontiers in cellular neuroscience},
  volume={13},
  pages={457375},
  year={2019},
  publisher={Frontiers},
  doi={10.3389/fncel.2019.00159}
}
\end{document}